\documentclass[twocolumn]{IEEEtran}

\ifCLASSINFOpdf
\else
\fi
\usepackage{dblfloatfix}    

\usepackage{cite}
\usepackage{amsmath,amssymb,amsfonts}
\usepackage{bbm}

\usepackage{graphicx}
\usepackage{textcomp}
\usepackage{xcolor}
\usepackage{colortbl} 

\usepackage{url} 

\usepackage{bm}
\usepackage{supertabular}
\usepackage{amsthm}
\theoremstyle{definition}

\usepackage[linesnumbered,ruled,vlined]{algorithm2e}
\usepackage{pbox}
\usepackage{float} 
\usepackage{subcaption} 

\usepackage{boldline,multirow}
\usepackage{array}

\usepackage{pbox}

\usepackage{caption}

\usepackage{dblfloatfix}

\usepackage[normalem]{ulem}

\usepackage{cancel} 

\usepackage{pifont}

\usepackage{tcolorbox}

\usepackage{enumitem}

\begin{document}


\title{Handoff Design in User-Centric Cell-Free Massive MIMO Networks Using DRL}
%
%
\author{Hussein~A.~Ammar\IEEEauthorrefmark{1},~\IEEEmembership{Member,~IEEE}, 
	Raviraj~Adve\IEEEauthorrefmark{2},~\IEEEmembership{Fellow,~IEEE},
	Shahram~Shahbazpanahi\IEEEauthorrefmark{3}\IEEEauthorrefmark{2},~\IEEEmembership{Senior Member,~IEEE},
	Gary~Boudreau\IEEEauthorrefmark{4},~\IEEEmembership{Senior Member,~IEEE}
	and~Israfil Bahceci\IEEEauthorrefmark{4}~\IEEEmembership{Member,~IEEE}
	\thanks{
		\IEEEauthorrefmark{1}H. A. Ammar is with the Department of Electrical and Computer Engineering (ECE), Royal Military College of Canada, Kingston, ON K7K 7B4, Canada (e-mail: hussein.ammar@rmc.ca).
	}
	\thanks{
		\IEEEauthorrefmark{2}R. Adve is with the Department of ECE, University of Toronto, Toronto, ON M5S 3G4, Canada (e-mail: rsadve@comm.utoronto.ca).
	}
	\thanks{
		\IEEEauthorrefmark{3}S. Shahbazpanahi is with the Department of Electrical, Computer, and Software Engineering, University of Ontario Institute of Technology, Oshawa, ON L1H 7K4, Canada. He also holds a Status-Only position with the Department of ECE, University of Toronto.
	}
	\thanks{
		\IEEEauthorrefmark{4}G. Boudreau and I. Bahceci are with Ericsson Canada, Ottawa, ON K2K 2V6, Canada.
	}
}


%
%

\maketitle 


\begin{abstract}
In the user-centric cell-free massive MIMO (UC-mMIMO) network scheme, user mobility necessitates updating the set of serving access points to maintain the user-centric clustering. Such updates are typically performed through handoff (HO) operations; however, frequent HOs lead to overheads associated with the allocation and release of resources. This paper presents a deep reinforcement learning (DRL)-based solution to predict and manage these connections for mobile users. Our solution employs the Soft Actor-Critic algorithm, with continuous action space representation, to train a deep neural network to serve as the HO policy. We present a novel proposition for a reward function that integrates a HO penalty in order to balance the attainable rate and the associated overhead related to HOs. We develop two variants of our system; the first one uses mobility direction-assisted (DA) observations that are based on the user movement pattern, while the second one uses history-assisted (HA) observations that are based on the history of the large-scale fading (LSF). Simulation results show that our DRL-based continuous action space approach is more scalable than discrete space counterpart, and that our derived HO policy automatically learns to gather HOs in specific time slots to minimize the overhead of initiating HOs. Our solution can also operate in real time with a response time less than $0.4~{\rm ms}$.

\end{abstract}

%
\IEEEpeerreviewmaketitle

%
%
%
\section{Introduction}
User-centric cell-free massive MIMO~(UC-mMIMO) is a wireless network architecture where each user is served by a custom group of \textit{neighboring} access points~(APs) which are connected to a central unit (CU) via fronthaul links~\cite{cellFreeVersusSmallCells7827017}. Unlike the current cellular system that is based on macro base stations, UC-mMIMO deploys cooperative APs that jointly serve users without relying on a traditional cellular boundaries.

UC-mMIMO helps to achieve reliable wireless connectivity and provides uniform performance throughout the network~\cite{cellFreeVersusSmallCells7827017, 8845768}. However, this beyond-5G mobile wireless network architecture introduces the key challenge of determining the connections between the APs and the users when moving through the network~\cite{surveyAmmar9650567}. 


Although user mobility plays a central role in the development of wireless networks, and hence the UC-mMIMO network scheme, it is often overlooked, and instead, network analysis and optimization mostly involve a static snapshot of the network~\cite{9615200, 9500543}. Mobility requires the user to update its set of serving APs, which is done through handoffs~(HOs). 
%
%
To keep serving the mobile users by neighboring APs and hence maintain the UC-mMIMO scheme, frequent HOs are needed~\cite{mobilityMMwaveCellFree9616361}. However, this leads to a high load on the control plane and a repeated reservation and release of communication resources, which will impact the network performance and scalability~\cite{surveyAmmar9650567}. Conversely, less frequent HOs result in serving the mobile user by a set of distant APs, compromising the user-centric scheme and affecting performance. 

Managing HOs in UC-mMIMO networks is more challenging than that in conventional cellular networks due to the absence of cells on the access channel and the simultaneous connections between each user and multiple APs. To address this, a user-APs \textit{association strategy}~\cite{zaher2022soft} is needed to limit unnecessary changes in the serving set of each user. In general, HO strategies that capitalize on the benefits of cell-free communications are essential for connection stability~\cite{beerten2023cell}. In particular, the HO strategies must consider a complex objective and the long-term performance rather than relying on simplistic HO decisions. In Table~~\ref{table:UC_vs_cellular}, we summarize these fundamental differences which makes managing HOs in UC-mMIMO more challenging than in cellular networks.

\begin{table}[t]
	\scriptsize
	\centering
	
	\caption{Handoff challenges in UC-mMIMO versus conventional cellular networks.}
	\vspace{-0.5em}
	\begin{tabular}{|p{0.22\linewidth}|p{0.31\linewidth}|p{0.31\linewidth}|}
			\hline
			\hline
			$\!\!\!$\textit{\textbf{Metric}} &
			\textit{\textbf{UC-mMIMO}}&
			$\!\!\!$\textit{\textbf{Conventional Cellular Network}}
			\\
			\hline
			User association & Connect to \textit{multiple APs} through a flexible user-centric approach & Connect to the \textit{single strongest BS}
			\\
			\hline
			Connection criteria & No strict cell boundaries & Predefined fixed cells
			\\
			\hline
			Protocol & Still not developed & Developed by 3GPP
			\\
			\hline
			HO trigger & Depend on multiple metrics & Crossing a cell boundary; bad signal quality
			\\
			\hline
			HO complexity & High 
			& Moderate
			\\
			\hline
			HO frequency & High; multiple serving APs
			& Low
			\\
			\hline
			Control overhead & High due to coordination 
			& Low 
			\\ 
			\hline
			Load balancing & More dynamic 
			& Less dynamic 
			\\
			\hline
			Channel Aging & More severe due to multiple user-AP connections, dense deployment, and coordination requirement between APs & Less severe as channel updates are needed at a single serving BS and due to lower deployment density for BSs
			\\
			\hline
			\hline
	\end{tabular}
	\label{table:UC_vs_cellular}   
	\vspace{-2.5em}
\end{table}


The topic of frequent HOs in UC-mMIMO networks is an intriguing area that has not been explored in detail. The authors in~\cite{mobilityMMwaveCellFree9616361} consider minimizing the number of HOs in UC-mMIMO network with millimeter wave communications, however the approach is heuristic and does not consider optimality in terms of data rate and connection stability. To the best of our knowledge, this is the first investigation that focuses on controlling the number of HOs in UC-mMIMO systems using both a sequential Markov decision process~(MDP) and deep reinforcement learning (DRL), where other studies such as~\cite{zaher2022soft} considered mobility and HOs aspect without obtaining an explicit HO policy or using the aforementioned tools. 

User mobility causes \textit{channel aging}~\cite{CSIAging6608213}, which is the discrepancy between the channel state information~(CSI) at the time of estimation and at the time for data transmission. This seems to affect the quality of the estimated CSI and the system performance~\cite{channelAgingMassiveMIMO8122014}. 
The studies in~\cite{CSIAgingzheng2020cell, 9838357} highlight the importance of considering mobility in the analysis of UC-mMIMO networks and study the effect of channel aging on performance. Specifically, the authors derive the uplink spectral efficiency under large-scale fading decoding and matched-filter receiver cooperation. The authors in~\cite{ChAgingPhaseNoise9471851} consider the effect of phase noise and channel aging on the zero-forcing beamforming scheme.

The authors in~\cite{9103233} and~\cite{cellFreeHardwareImpair} consider channel aging with hardware impairments assuming Rayleigh and Rician fading channels, respectively. In their research paper~\cite{9416909}, the authors investigate 
methods to mitigate the effect of channel aging, including performing statistical channel cooperation power control and adjusting the length of the resource block.

Overall, the discourse concerning mobility in UC-mMIMO systems has focused on investigating the effect of channel aging and developing methods to mitigate it, rather than on optimizing the HO decisions. It is precisely this research gap that our study aims to address.

\begin{table}[t]
	\scriptsize
	\centering
	\caption{Classification for different references on handoff designs; \checkmark$\!\!\!\backslash$ means partially considered.}
	\vspace{-0.5em}
	\begin{tabular}{|p{0.06\linewidth}|p{0.06\linewidth}|p{0.06\linewidth}|p{0.02\linewidth}|p{0.07\linewidth}|p{0.05\linewidth}|p{0.05\linewidth}|p{0.02\linewidth}|p{0.08\linewidth}|p{0.03\linewidth}|}
		\hline
		\hline
		$\!\!\!$\textit{\textbf{Ref.}} &
		\tiny\textit{\textbf{UC-mMIMO}}&
		\tiny 
		$\!\!\!$\textit{\textbf{Mobility effects}} 
		& 
		$\!\!$\tiny 
		\textit{\textbf{DRL}}
		&
		\tiny 
		$\!\!$\textit{\textbf{Continuous action space}} 
		&
		$\!\!$\tiny 
		\textit{\textbf{Explicit HO policy}} 
		&
		$\!\!\!$\tiny 
		\textit{\textbf{Minimize HOs}} 
		&
		$\!\!\!$\tiny 
		\textit{\textbf{MDP}}
		&
		\tiny 
		\textit{\textbf{Partially observed state}}	
		&
		\tiny 
		\textit{\textbf{Year}} 
		\\
		\hline
		\cite{mobilityMMwaveCellFree9616361} & \checkmark & \ding{55} & \ding{55} & \ding{55} & \checkmark & \ding{55} & \ding{55}
		& \ding{55} &$\!\!\!$2021
		\\
		\hline
		\cite{zaher2022soft}
		& \checkmark & \ding{55} & \ding{55} & \ding{55} & \ding{55} & \checkmark$\!\!\!\backslash$ & \ding{55}
		& \ding{55} &$\!\!\!$2024
		\\
		\hline
		\cite{9275345}& 
		\checkmark$\!\!\!\backslash$
		& \ding{55} & \checkmark & \checkmark & \checkmark & \checkmark & \checkmark & \ding{55} &$\!\!\!$2020
		\\
		\hline
		\cite{8834857}& \ding{55} & \ding{55} & \checkmark 
		& \ding{55} & \checkmark & \ding{55} &
		\checkmark & \ding{55} &$\!\!\!$2019
		\\
		\hline
		\cite{8387430} & \ding{55} & \ding{55} & \checkmark 
		& \checkmark & \checkmark & \checkmark & \checkmark
		& \ding{55} & $\!\!\!$2018
		\\
		\hline
		\cite{POMDP_J} & \checkmark & \checkmark & \ding{55} & \ding{55} & \ding{55} & \checkmark & \checkmark
		& \checkmark &$\!\!\!$2024
		\\
		\hline
		$\!\!\!$Current work & \checkmark & \checkmark & \checkmark & \checkmark & \checkmark
		& \checkmark
		& \checkmark
		& \checkmark
		&
		\\
		\hline
		\hline
	\end{tabular}
	\label{table:ref_compare}   
	\vspace{-2em}
\end{table}

Due to its significant capabilities, machine learning has been utilized to manage HOs in conventional cellular networks through various methods. The authors in~\cite{alkhateeb2018machine} use gated recurrent neural networks to predict and prevent blockages in millimeter wave (mmWave) communications. Basically, the base stations~(BSs) utilize past beamformer sequences to forecast the blockage probability of the links in the upcoming time frames. Then, proactive HO decisions are performed through connecting the user to a BS with a high probability of a line-of-sight link. 
The authors in~\cite{8648402} propose a HO management scheme for wireless local area networks based on DRL using deep Q-networks~(DQNs); however, the wireless aspects resulting from mobility are not considered.


The authors in~\cite{8067509} derive an RL-based HO policy that reduces the number of HOs while maintaining user quality of service~(QoS) requirements for mmWave heterogeneous networks. Their approach comprises two parts; one that triggers HOs using the mmWave channel characteristics and QoS requirements, and the other that selects the target BS. Here, RL is based on a tabular policy approach.
These aforementioned methods assume that a single BS serves the users, and hence, they do not explore HOs for the UC-mMIMO network configuration, where the user is served by a group of neighboring APs that are not bounded by cells. Therefore, they address a fundamentally different problem from our study.

Studying HOs is different in nature from constructing the users' serving sets in a static UC-mMIMO network~\cite{ammar9519163, ammar9570126, 10517887, 9839273}. An efficient solution for HOs requires considering the temporal sequence of events and the relation between them, this often can be modeled as a Markov process~\cite{conf_POMDP}. By adding a feedback controller, we obtain a MDP, which comprises the formal model to define a RL environment.

Motivated by this consideration, we model the task of HO management in the UC-mMIMO network scheme as an MDP, and we construct a DRL framework to develop our HO solution. Specifically, we use the soft actor-critic~(SAC) algorithm~\cite{haarnoja2018soft2} with a \textit{continuous action space representation} to train an actor DNN to take HO decisions. The term ``actor-critic'' refers to a structure that incorporates both policy-based and value-based DRL methods. The actor part refers to a DNN representing the policy which is used to determine the actions. This DNN is then used to select the actions even when the DRL training is done. The critic part refers to a DNN that criticizes the actor by evaluating its corresponding actions and hence helping to guide the updates for the actor DNN. It is worth mentioning that the actor-critic method is one of the most powerful methods in DRL. The training can be performed offline to optimize the HO decisions and maximize system performance. Once the DNN is trained, it can be used online, and with real-time response, as the HO policy for the UC-mMIMO network. Our approach surmounts the high computational complexity and the limited scalability of tabular policies obtained using value iteration algorithms~\cite{POMDP_J, conf_POMDP}.
%


We propose two variants for our system, each uses distinct information to anticipate the quality of the large-scale fading (LSF) between the user and each AP in the future, hence, aiding in optimizing HO decisions. The first variant uses mobility direction-assisted observations, in which our DNN utilizes the user's direction of motion. While the second variant uses history-assisted observations, where our DNN utilizes the history of the LSF statistics (past samples) between the user and each AP. 
We use a penalized reward function that incorporates the achievable rate with the effect of channel aging and channel estimation error included. The penalty represents a weighted overhead due to HOs, where the HO penalty is based on a non-linear function that can assign costs for initiating HOs.

HO decisions are necessarily discrete, however, this leads to an action space that grows exponentially in size as a function of the  number of APs. Utilizing a continuous action space in our DRL framework is, therefore, more scalable than using a discrete action space. Our results also show that the DNN can be trained to take optimized HO decisions for the moving user, hence serving as a DRL-based HO policy. Furthermore, the results demonstrate that this policy can outperform the state-of-the-art LSF-based HO scheme when the penalty for initiating HOs is high. In general, the solution with movement direction-assisted observation provides better performance compared to the history-assisted observation variant. Moreover, results show that our HO policy can provide real-time decisions even for large network sizes. We also evaluate the effectiveness of our solution under partial observability by representing our problem using a partially observable MDP (POMDP), where our solution stays efficient in managing HOs.


The contributions of our study can be summarized as:
\begin{itemize}
	\item We develop a DRL-based solution to manage HOs in UC-mMIMO. To train a DNN to serve as our HO-policy, we use the SAC algorithm with a continuous action space representation and a reward function that penalizes HOs. To the best of our knowledge, our work is the first to employ DRL for this purpose in UC-mMIMO network.
	\item We introduce two variants for our system: the first is a movement direction-assisted solution, and the second is a history-assisted solution. Both systems leverage their own information to aid in HO decision-making and thus minimize the number of HOs required as the user moves. In comparison to LSF-based HOs, our DRL-based HO policy obtained for the movement direction-assisted and history-assisted variants can improve the achievable rate by approximately $27\%$ and $20\%$, respectively.
	\item We design our DNN to output a vector of continuous action space, which is then processed through a mapping function to determine the user's connection decisions. Although HO decisions are discrete, which might suggest a natural choice of a discrete action space, we show that our approach, which utilizes a vector of continuous action space, is more scalable and reduces the policy response time by $75\%$ and the disk size by $98\%$ for large network sizes (Table~\ref{table:diskSpace}). Additionally, our approach supports real-time control with a latency of under $0.4~{\rm ms}$, i.e., less than the duration of a 5G new radio (NR) subframe, making it suitable for online applications.
	\item We evaluate the effectiveness of our solution under the more realistic case of partial observability by representing our problem as a POMDP. Results indicate that our solution is efficient, where the loss in the achievable rate is less than~$10\%$ compared to the fully observable case of an MDP. 
	This outcome underscores the potential of DRL to deploy HO policies in imperfect, real-world, situations.
\end{itemize}

The rest of the paper is organized as follows. Section~\ref{section:Model} defines our system model including the dynamics required to characterize the effect of mobility on the communication. Section~\ref{section:RL_model} frames the problem as an RL model with its formal MDP explained. Section~\ref{sec:DRL} describes how we employ our DRL framework using SAC to train a DNN to function as the HO policy. Section~\ref{sec:POMDP} details defining the problem as a POMDP instead of an MDP which is a necessary step for practical deployment scenarios. Section~\ref{sec:results} presents simulation results and insights. Section~\ref{sec:conclusion} summarizes our conclusion.

Both lower and upper case letters represent scalars, while their bold counterparts, e.g., ${\bf a}$ and ${\bf A}$, represent vectors and matrices, respectively. Operators $(\cdot)^T$, $(\cdot)^*$, $(\cdot)^H$, and $\mathbb{E}\{\cdot\}$, denote the transpose, conjugate, and conjugate transpose, and statistical expectation, respectively. Symbols $\|\cdot\|_2$ and $|\cdot|$ are the vector and scalar Euclidean norms, and ${\bf I}_m$ is $m \times m$ identity matrix. For a set $\mathcal{A}$, $|\mathcal{A}|$ is its cardinality. Symbols $\mathbb{B}$, $\mathbb{R}$, and $\mathbb{C}$ represent binary, real, and complex numbers, respectively. ${\bm x} \sim \mathcal{CN}({\bm \mu}, {\bf R})$ denotes a complex Gaussian random variable with mean ${\bm \mu}$ and covariance matrix ${\bf R}$.

\vspace{-0.5em}
\section{System Model}\label{section:Model}
\vspace{-0.4em}
\subsection{Network model}
\vspace{-0.2em}

\begin{figure}[t]
	\centering
	\includegraphics[width=0.95\columnwidth]{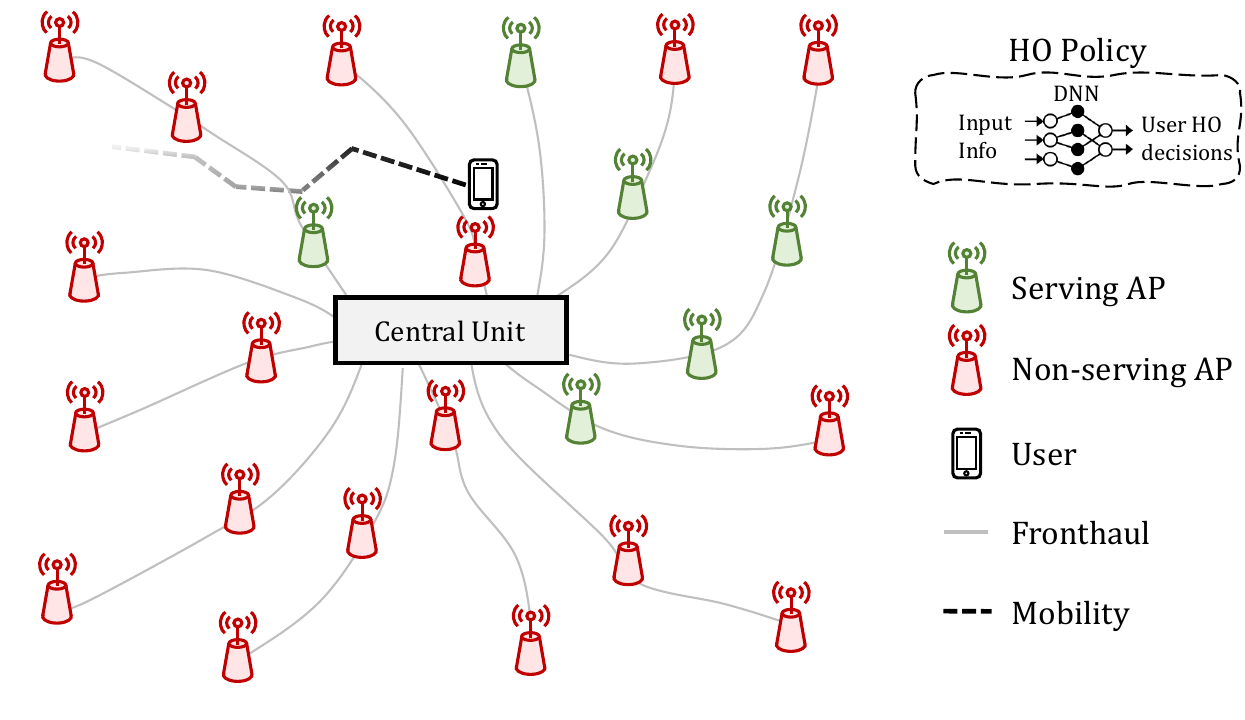}
	\vspace{-1.2em}
	\caption{UC-mMIMO network depicting a moving user and a setup for managing HO decisions.}
	\label{fig:UserCentricClustering}
	\vspace{-1.5em}
\end{figure}

We consider a network comprising $B$ APs represented by the set $\mathcal{B}$ and connected to a central unit (CU) via wired fronthaul connections, such as the radio stripes system~\cite{frenger2019antenna}. The APs, each equipped with $M$ antennas, jointly serve the users represented by the set $\mathcal{U}$ using the UC-mMIMO scheme. Specifically, each user is served by a set of neighboring APs, and the concept of cells is not applicable to the access channel. We assume a~$\lambda/2$ antenna spacing and a rich scattering environment that would make the spatial correlation for the small-scale fading across each AP antennas negligible. For the large-scale fading, the details will be provided later in Section~\ref{section:spatialCorreLSF}.

We assume that communication is segmented into discrete decision time steps. During each decision step $t$, a typical user $u \in \mathcal{U}$ is served by several APs simultaneously through coherent transmissions~\cite{ammar9519163}. The set of APs that serve user $u$ during decision step $t$ is denoted as the serving set $\mathcal{C}_u^{(t)}$. As user $u$ moves, $\mathcal{C}_u^{(t)}$ is updated by adding and removing APs as needed by initiating HOs. Hence, HOs occur at the beginning of each decision step $t$ if $\mathcal{C}_u^{(t)} \ne \mathcal{C}_u^{(t-1)}$. From the APs's side, the users to be served by AP $b$ are represented by the set $\mathcal{E}_b^{(t)}$.

Fig.~\ref{fig:UserCentricClustering} depicts the network, where we focus on a typical user $u$ moving using any mobility pattern, e.g., a straight line, or the random waypoint mobility model~\cite{ammar_RWP9531946}. 
This figure shows a trained DNN that acts as the HO policy by managing the connection decisions of each user in the network independently. It is worth mentioning that in the 3GPP Release~16~\cite{3GPPTS38.300}, user HO decisions are made independently based on mobile-assisted signal measurements and radio resource management information, which aligns with the fundamental concept employed in this paper.
%

\vspace{-1em}
\subsection{Channel model}
%
\subsubsection{Channel Aging Model}
User movement causes temporal variations in the wireless channels involved even within a single communication cycle that includes a channel estimation phase and a data transmission phase. Consequently, when a user moves at a relatively high speed, utilizing a block fading channel model with a specific coherence time is inaccurate; instead, a time-varying channel model is necessary. This is why we use the term ``communication cycle'' instead of the conventional term ``channel coherence interval''.

A communication cycle contains $\tau_{\rm c}$ channel uses comprising an uplink pilot training phase of length $\tau_{\rm p}$ and a downlink data transmission phase of length $\tau_{\rm d}$. Within a communication cycle, the small-scale fading part of the channel is considered to be a wide-sense stationary (WSS) process~\cite{CSIAging6608213}, while the large-scale fading is considered constant because it changes at a slower rate compared to the duration $\tau_{\rm c}$.

Specifically, the channel realization between AP $b$ and user $u$ at channel use instant $n = 0,\dots,\tau_{\rm c}-1$ is modeled as ${\bf h}_{bu}[n] \triangleq \sqrt{ \beta_{bu} } {\bf g}_{bu}[n] \in \mathbb{C}^{M \times 1}$, where ${\bf g}_{bu}[n] \sim \mathcal{CN}\left({\bf 0}, {\bf I}_M \right)$ is the small-scale fading, and $\beta_{bu}$ is the large-scale fading that accounts for the shadowing and path loss. 
The channel is temporally correlated, i.e., for any channel uses $n$ and $n'$, ${\bf g}_{bu}[n]$ and ${\bf g}_{bu}[n']$ are correlated~\cite{CSIAgingzheng2020cell}. The temporal correlation of the channel at $n$ and $n'$; $n, n'=0, \dots,\tau_{\rm c}-1$, can be characterized using Jakes' model~\cite{jakes1994microwave} 
through the temporal correlation coefficient $\rho_{u}[n - n']$ defined~as
\begin{align}\label{eq:aging}
	\rho_{u}[n - n'] \triangleq J_0 \left( 2 \pi \left( n - n' \right) f_{{\rm{D}}_u} T_{\rm s} \right)
\end{align}
where 
$J_0(\cdot)$ is the zeroth-order Bessel function of the first kind, $T_{\rm s}$ is the sampling period of each channel use, and $f_{{\rm{D}}_u} = \frac{f_{\rm c}  \mathtt{v}_u}{\rm c}$ is the maximum Doppler shift which depends on the mobility speed $\mathtt{v}_u$ of the user, the carrier frequency $f_{\rm c}$, and the speed of light ${\rm c}$. 

Based on~\eqref{eq:aging}, for each communication cycle of length $\tau_{\rm c}$, the small-scale fading at instant $n$ can be written as a function of initial state ${\bf g}_{bu}[0]$ and an innovation component ${\bf v}_{bu}[n]$ as~\cite{channelAgingMassiveMIMO8122014}
\begin{align}\label{eq:channelEvolution}
	{\bf g}_{bu}[n] = \rho_{u}[n] {\bf g}_{bu}[0] + \bar{\rho}_{u}[n] {\bf v}_{bu}[n],
\end{align}
where ${\bf v}_{bu}[n] \sim \mathcal{CN}\left({\bf 0}, {\bf I}_M \right)$ is the innovation component at instant $n$ and it independent of the small-scale fading, $\rho_{u}[n]$ is the temporal correlation coefficient of user $u$ between channel realizations at instants $0$ and $n$, with $0 \le \rho_{u}[n] \le 1$, and $\bar{\rho}_{u}[n] = \sqrt{ 1 - |\rho_{u}[n]|^2 }$.

\subsubsection{Channel Estimation}
Due to user mobility, aging affects the channel instants in both the data and pilot transmission phases. To construct a proper signal model, we need to relate the channel at the instant it was estimated, which we denote as instant $n_{\rm est}$, to the other channel use instants.

As is common when analyzing channel aging, we use the delta-function for the channel training sequence. We define the set $\mathcal{U}_{i}$ which corresponds to the users using the pilot signal $\delta[n - i]$, which means a user $u \in \mathcal{U}_{i}$ uses the pilot signal $\phi_u[n] = \delta[n - i]$, with $\phi_u^H[n] \phi_u[n] = 1$ and $1 \le i \le \tau_{\rm p}$, where $\tau_{\rm p}$ is the length of the pilot transmission phase. Here, $\delta[n] = 0$ for $n \ne 0$ and $\delta[n] = 1$ for $n = 0$. 
At instant $i$ during the uplink pilot training phase, i.e., $i \le \tau_{\rm p}$, the signal received at AP $b$ can be written as
\begin{align}\label{eq:Signal_at_instant_i}
	{\bf y}_{b}[i]
	&= \sum_{u' \in \mathcal{U}_i} \sqrt{p^{({\rm u})}} 
	{\bf h}_{bu'}[i]
	+ 
	{\bf z}_{b}[i]
\end{align}
where $p^{({\rm u})}$ is the uplink transmit power of each user, and ${\bf z}_{b} \sim \mathcal{CN}({\bf 0}, \sigma_{\rm z}^2 {\bf I}_M)$.

Channel estimation occurs after all the training signals are received, i.e., at $n_{\rm est} = \tau_{\rm p} + 1$. Based on~\eqref{eq:channelEvolution}, ${\bf g}_{bu}[n]$ at time instant $n \le \tau_{\rm p}$ can be related to ${\bf g}_{bu}[n_{\rm est}]$ as follows:
\begin{align}\label{eq:channelEvolutionChanEst}
	{\bf g}_{bu}[n] = \rho_{u}[n_{\rm est} - n] {\bf g}_{bu}[n_{\rm est}] + \bar{\rho}_{u}[n_{\rm est} - n] {\bf v}_{bu}[n]
\end{align}

Plugging~\eqref{eq:channelEvolutionChanEst} in the small-scale fading component in the channel ${\bf h}_{bu}[i]$ in~\eqref{eq:Signal_at_instant_i}, and assuming that AP $b$ is estimating the channel for user $u \in \mathcal{U}_i$, the signal in~\eqref{eq:Signal_at_instant_i} received at AP~$b$ at instant~$i$ during the uplink pilot training phase is
\begin{align}\label{eq:Signal_at_instant_i_detailed}
	{\bf y}_{b}[i]
	&=
	\resizebox{0.41\columnwidth}{!}
	{$
	\sqrt{p^{({\rm u})}} 
	\rho_{u}[n_{\rm est} - i]
	{\bf h}_{bu}[n_{\rm est}] 
	$}
	+
	\resizebox{0.41\columnwidth}{!}
	{$
	\sqrt{p^{({\rm u})} \beta_{bu}}
	\bar{\rho}_{u}[n_{\rm est} - i] {\bf v}_{bu}[i]
	$}
	\nonumber \\
	&\quad
	+ 
	\resizebox{0.38\columnwidth}{!}
	{$
	\sum_{u' \in \mathcal{U}_i, u' \ne u} \sqrt{p^{({\rm u})}} 
	{\bf h}_{bu'}[i]
	$}
	+ 
	{\bf z}_{b}[i]
\end{align}
%
Using linear minimum mean square error (MMSE), AP $b$ estimates the channel of user $u$ as
\begin{align}\label{eq:est_chan}
	{\bf \hat{h}}_{bu}[n_{\rm est}]
	=
	\frac{ \rho_{u}[n_{\rm est} - i ] \sqrt{p^{({\rm u})}} \beta_{bu}}{ \sum_{u' \in \mathcal{U}_i} p^{({\rm u})} \beta_{bu'} + \sigma_{\rm z}^2 }
	{\bf y}_{b}[i],
	\quad \text{for}\ u \in \mathcal{U}_i
\end{align}
When estimating the channel using linear MMSE, the channel estimation error $\mathrm{ {\bf e}}_{bu} = {\bf h}_{bu}[n_{\rm est}] - {\bf \hat{h}}_{bu}[n_{\rm est}]$ is uncorrelated with the estimated channel ${\bf \hat{h}}_{bu}[n_{\rm est}]$ and is distributed as $\mathrm{ {\bf e}}_{bu} \sim \mathcal{CN}\left({\bf 0}, {\bm \Theta_{bu}}\right)$, where the covariance ${\bm \Theta_{bu}} \triangleq \beta_{bu} {\bf I}_M - \psi_{bu} {\bf I}_M$, where $\psi_{bu}$ is defined as
\begin{align}\label{eq:var_estCh}
	\psi_{bu}
	=
	\frac{\rho_{u}^2[ n_{\rm est} - i  ] p^{({\rm u})} \beta_{bu}^2 }{ \sum_{u' \in \mathcal{U}_i} p^{({\rm u})} \beta_{bu'} + \sigma_{\rm z}^2 },
	\quad \text{for}\ u \in \mathcal{U}_i
\end{align}

\subsection{Downlink Signal Model and Achievable Rate}
The APs utilize conjugate beamforming to serve the users, which is optimal in the single user case we consider for initiating the HO decisions. As noted earlier, the joint processing of users for HO decisions is outside the scope of this study.

Although conjugate beamforming may not be the optimal beamformer choice, it is often selected in UC-mMIMO network schemes for its mathematical tractability, which allows for greater focus on the core problem under investigation. Additionally, the usage of conjugate beamforming in this context has zero overhead on the exchange of CSI and the control plane between the APs and the CU.

Through conjugate beamforming, AP $b$ uses the channels estimated at time instant $n_{\rm est}$ to transmit data to the users. Based on this, the signal ${\bf x}_b[n] \in \mathbb{C}^{M \times 1}$ sent by AP $b$ at time instant $n \ge n_{\rm est}$ to users $\mathcal{E}_b$ can be written as
\begin{align}
	{\bf x}_b[n] =
	\sum_{u \in \mathcal{E}_b}
	\sqrt{\eta_{bu}}
	{\bf \hat{h}}_{bu}^*[n_{\rm est}] s_{u}[n],
\end{align}
where ${\bf \hat{h}}_{bu}^*[n_{\rm est}]$ denotes the conjugate of the channel estimated at time instant $n_{\rm est}$, $s_{u}$ is the complex data symbol for the user with $\mathbb{E}\{|s_{u}[n]|^2\} = 1$, and $\eta_{bu}$ is a \textit{statistical} normalizing term~\cite{9141340} for the transmit power allocated by AP $b$ to user $u$. This term allows the AP to satisfy, on-average, the power budget $p^{({\rm d})}$ available, i.e., $\mathbb{E}\{ \| {\bf x}_b[n]\|^2 \} \le p^{({\rm d})}$.

Channel aging affects even the data transmission phase. Using~\eqref{eq:channelEvolution} for the channel evolution, the small-scale fading at time instant $n \ge n_{\rm est}$ can be also represented as
\begin{align}\label{eq:channelEvolution_data}
	{\bf g}_{bu}[n] = \rho_{u}[n - n_{\rm est}] {\bf g}_{bu}[n_{\rm est}] + \bar{\rho}_{u}[n - n_{\rm est}] {\bf v}_{bu}[n]
\end{align}

By employing~\eqref{eq:channelEvolution_data} and our analysis, the signal received at user $u$ during the downlink data transmission phase, i.e., at $n \ge n_{\rm est}$, can be written as~\eqref{eq:signalModel} shown on page~\pageref{eq:signalModel}, 
\begin{figure*}[b]
	\hrule
\begin{align}\label{eq:signalModel}
	&y_{u}[n] =
	\sum_{b \in \mathcal{B}} 
	{\bf h}_{bu}^T[n] {\bf x}_b[n]
	+ z_u
	\nonumber \\
	&\quad =
	\sum_{b \in \mathcal{C}_u} \sqrt{\eta_{bu} \beta_{bu}} 
	\left(
	\rho_{u}[n - n_{\rm est}] {\bf g}_{bu}^T[n_{\rm est}] + \bar{\rho}_{u}[n - n_{\rm est}] {\bf v}_{bu}^T[n]
	\right)	
	{\bf \hat{h}}_{bu}^*[n_{\rm est}]
	s_{u}[n]
	+
	\sum_{u' \in \mathcal{U}, u' \ne u}
	a_{uu'}[n] s_{u'}[n]
	+ z_u
	\nonumber \\
	& \quad =
	\underbrace{
	\rho_{u}[n - n_{\rm est}]
	\sum_{b \in \mathcal{C}_u} \sqrt{\eta_{bu}} 
	\mathbb{E}\left\{
	{\bf h}_{bu}^T[n_{\rm est}]
	{\bf \hat{h}}_{bu}^*[n_{\rm est}]
	\right\}
	s_{u}[n]
	}_{\mathrm{desired~signal}~{\rm DS}_u}
	+
	\underbrace{
		\rho_{u}[n - n_{\rm est}]
		\sum_{b \in \mathcal{C}_u} \sqrt{\eta_{bu}}  
		\left(
		{\bf h}_{bu}^T[n_{\rm est}]
		{\bf \hat{h}}_{bu}^*
		-
		\mathbb{E}\left\{
		{\bf h}_{bu}^T[n_{\rm est}]
		{\bf \hat{h}}_{bu}^*
		\right\}
		\right)
		s_{u}[n]
	}_{\mathrm{beamformer~uncertainity}~{\rm BU}_u}
	\nonumber \\[-5pt]
	& \quad \quad
	+
	\underbrace{
	\bar{\rho}_{u}[n - n_{\rm est}]
	\sum_{b \in \mathcal{C}_u} \sqrt{\eta_{bu} \beta_{bu}} 
	{\bf v}_{bu}^T[n]
	{\bf \hat{h}}_{bu}^*
	s_{u}[n]
	}_{\mathrm{channel~aging}~{\rm CA}_u}	
	+
	\sum_{u' \in \mathcal{U}, u' \ne u}
	\underbrace{
		a_{uu'}[n] s_{u'}[n]
	}_{\mathrm{multiuser~interference~{\rm MI}_u}}
	+ \underbrace{z_{u}}_{\mathrm{noise}},
\end{align}
\vspace{-2em}
\end{figure*}
where $a_{uu'}[n] = \sum_{b' \in \mathcal{C}_{u'}} \sqrt{\eta_{b'u'}} {\bf h}_{b'u}^T[n] {\bf \hat{h}}_{b'u'}^*$.

Using the signal model in~\eqref{eq:signalModel}, 
we characterize the rate performance through a lower bound for the channel capacity~as
\begin{align}\label{eq:rate_LB}
	R_u^{(\rm lb)} &=
	\frac{1}{\tau_{\rm c}} \sum_{n=n_{\rm est}}^{\tau_{\rm c}} \log \left( 1 + \frac{ \left| \mathbb{E}\left\{{\rm DS}_u[n]\right\} \right|^2 }{ A_u [n] } \right)
\end{align}
with
\begin{align}
	A_u [n] &=
	\mathbb{E}\left\{\left|{\rm BU}_u[n]\right|^2\right\}
	+ \mathbb{E}\left\{\left| {\rm CA}_u[n] \right|^2\right\}
	\nonumber \\[-2pt]
	&\quad
	+ \sum_{u' \in \mathcal{U}, u' \ne u}
	\mathbb{E}\left\{\left| {\rm MI}_{uu'}[n] \right|^2\right\}
	+ \sigma_{\rm z}^2
	\nonumber \\[-2pt]
	&=
	\xi_{2,3,u}[n] + \sum_{u' \in \mathcal{U}, u' \ne u} \xi_{4,uu'}[n] + \sigma_{\rm z}^2
\end{align}
where 
the power of the desired signal, the power of the signal which corresponds to the beamformer uncertainty + channel aging, and the power of the interference can be, respectively, written in closed-form expressions~as
{\allowdisplaybreaks
\begin{align}
	\xi_{1,u}[n] & =
	M^2
	\rho_{u}^2[n - n_{\rm est}]
	\Big|
	\sum_{b \in \mathcal{C}_u^{(t)}}
	\sqrt{\eta_{bu}} 
	\psi_{bu}
	\Big|^2
	%
	%
	\\[-5pt]
	\xi_{2,3,u}[n] & =
	M^2
	\sum_{b \in \mathcal{C}_u^{(t)}}
	\eta_{bu} \beta_{bu} 
	\psi_{bu}
	%
	%
	\\[-5pt]
	\xi_{4,uu'}[n] & =
	\nonumber \\
	&
	\hspace{-3em} 
	\begin{cases}
		\begin{aligned}
			&
			M^2 \bigg(\sum_{b' \in \mathcal{C}_{u'}^{(t)}} \eta_{b'u'}
			\beta_{b'u}\psi_{b'u'}
			\\[-5pt]
			&
			+
			\rho_{u}^2[n - n_{\rm est}]
			\bigg|\sum_{b' \in \mathcal{C}_{u'}^{(t)}} \sqrt{\eta_{b'u'}} \sqrt{\psi_{b'u} \psi_{b'u'} }
			\bigg|^2
			\bigg)
		\end{aligned}
		,& \text{if}\ u' \in \mathcal{U}_i
		\\
		M^2 \sum_{b' \in \mathcal{C}_{u'}^{(t)}} \eta_{b'u'}
		\beta_{b'u}\psi_{b'u'}
		,& \text{if}\ u' \notin \mathcal{U}_i
	\end{cases}
\end{align}}
%
%
%
%
$\!\!$The derivation for~\eqref{eq:rate_LB} is shown in Appendix~\ref{Appendix:rate_LB}.
%
%
%
%

\section{Reinforcement Learning Model}\label{section:RL_model}
In this section, we describe the RL environment that models our HO problem. By constructing a reward-based system that sets a weighted penalty for HOs, we can tune the HO behavior of the user allowing to collect HOs in specific decision cycles only. It is worth emphasizing that the time between HO decisions may cover multiple communication cycles, hence HO decisions are not based on small-scale fading.

%
Our RL scheme is a self-learning approach composed of an environment that models the UC-mMIMO network, and an agent that interacts with this environment through time steps. In each step, an RL agent receives an observation that describes the state of the environment and executes an action to receive a reward. The RL environment can be formally described through an MDP where the state is fully observable.

\subsection{Actions and Action Space}\label{sec:ActionSpace}
At any decision step, each user of concern, which we call the typical user, will be served by a set of APs based on a HO policy. We use $a_{b}^{(t)} = 1$ to denote that the user is served by AP $b$ in decision step $t$ and $a_{b}^{(t)} = 0$ otherwise. Using this definition, the action at decision step $t$ is represented through the binary vector ${\bf a}^{(t)} = [a_{1}^{(t)}, \dots, a_{B}^{(t)} ]^T \in \mathbb{B}^{B \times 1}$. Hence, the serving set $\mathcal{C}_u^{(t)}$ represents the indices of the non-zero elements in ${\bf a}^{(t)}$. In order to implement the user-centric scheme, we select a particular size of $B_{\rm con} = |\mathcal{C}_u^{(t)}|$. This size is typically determined based on the QoS requirements of the user and the overall network load status. Hence, we assume that the user is served by $B_{\rm con} < B$ APs at each decision step.

%

\subsection{Observation Space}
The observation represents the input for our DNN. For a fully observable MDP, it is the same as the state of the system. 
We define our observation through the vector ${\bf o}^{(t)} \in \mathbb{R}^{4 B \times 1}$ defined~as
\begin{align}\label{eq:observation_state}
	{\bf o}^{(t)} &= \left[ {\rm S}\{[\log(\beta_{1u}^{(t)}) \ \dots \ \log(\beta_{Bu}^{(t)} ) ]\} \ {\rm S}\{[|\mathcal{E}_1| \ \dots \ |\mathcal{E}_B| ]\}
	\right.
	\nonumber \\[-5pt]
	& \quad\quad
	\left.
	{\rm S}\{[ a_{1u}^{(t-1)} \ \dots \ a_{Bu}^{(t-1)} ]\} 
	\ {\rm S}\{[ \zeta_{1u}^{(t)} \ \dots \ \zeta_{Bu}^{(t)} ]\} \right]^T
\end{align}
where $u$ represents the typical user for which the HO policy is derived, $\beta_{bu}^{(t)}$ is the LSF between the user and AP $b$, $|\mathcal{E}_b|$ is the number of user served by AP $b$, $a_{bu}^{(t-1)}$ is the connection decision between the user and AP $b$ at the previous decision step, and $\zeta_{bu}^{(t)}$ is a metric which gives information about the possibility for the channel between the user and the AP to be in a good state.

In preparation for the DRL framework, the log operation for the LSF statistics is used to limit the dynamic range of these values. The operator ${\rm S}\{\cdot\}$ applies a min-max normalization, then shifts and scales the input to make it between -1 and 1, hence, any entry in the observation vector ${\bf o}^{(t)}$ lies in $[-1, 1]$. This operation can be applied on the elements of a vector ${\bf x} = [\dots x_i \dots]^T$ through $x_i^{\rm scaled} = 2 \left( \frac{x_i - \min\{\bf x\}}{\max\{\bf x\} - \min\{\bf x\}} - 0.5 \right) $, which leads to a scaled version ${\bf x}^{\rm scaled}= [\dots x_i^{\rm scaled} \dots ]^T$. Our investigations show that this leads to an easier configuration for the hyperparameters of the DRL algorithm, which helps in making the training smoother and faster. It is worth noting that the observation vector includes the information of the LSF to all the APs in the network. We will consider partial observability later.

The aim behind using the parameters $\{\zeta_{bu}^{(t)}: b \in \mathcal{B}\}$ used in the observation vector is to prioritize the APs that are more likely to have a good channel connection to the user. This approach reduces unnecessary HOs and facilitates predictive HOs. We define two variants of our system to calculate $\zeta_{bu}^{(t)}$ denoted as follows:
\begin{align}
\zeta_{bu}^{(t)}
=
\begin{cases}
	\zeta_{bu}^{(t), ({\rm DA})},& \text{using DA observation} 
	\\
	\zeta_{bu}^{(t), ({\rm HA})},& \text{using HA observation} 
\end{cases}
\end{align}

Our first proposed variant uses the direction of the user's mobility, denoted as ``movement direction-assisted observations'', while our second variant uses the history of the state of the channel and is denoted as ``history-assisted observations''. As mentioned earlier, the information used by these two variants will help the DRL to learn which APs to prioritize for HO decisions, where the selected APs will be more likely to provide a good channel for the user. Hence, the parameters $\{\zeta_{bu}^{(t)}: b \in \mathcal{B}\}$ play a crucial role in deriving our HO policy.

\subsubsection{Movement Direction-Assisted Observations (DA)}
In using this variant, we assume that the direction of the movement of the user is known, which is a feasible, especially in beyond 5G networks, where accurate user location can be obtained. Moreover, the locations of the APs are known, because they are fixed. Hence, using this information, we can calculate the angle $\theta_{bu}$ between the direction of the user's movement and the direction of the line connecting the user to AP $b$. With this information, we determine $0 \le \zeta_{bu}^{(t), ({\rm DA})} \le 1$~as
\begin{align}\label{eq:zeta_direction}
	\zeta_{bu}^{(t), ({\rm DA})}
	=
	\frac{\cos(\theta_{bu}) + 1}{2}
\end{align} 

\subsubsection{History-Assisted Observations (HA)}
In using this variant, we leverage the history of the LSF between the user and the APs to gain insights on the potential quality of the channel in the next decision step. Specifically, the LSF between the user and AP $b$ at any decision step $t$ is classified as either good or bad based on the value of the LSF $\beta_{bu}^{(t)}$ compared to some threshold $\beta_{\rm threshold}$. Then, for each AP $b$ we calculate the probability of observing a good channel $0 \le \zeta_{bu}^{(t), ({\rm HA})} \le 1$ at decision step $t$ as
\begin{align}\label{eq:zeta_history}
	\zeta_{bu}^{(t), ({\rm HA})}
	\!\!\!
	=
	\begin{cases}
		\displaystyle
		\frac{
			\sum_{m = 1}^{t}
			\gamma_{\rm o}^{t-m}
			\mathbbm{1}\{
			\beta_{bu}^{(m-1)} > \beta_{\rm threshold}
			\}
		}{
			\sum_{m = 1}^{t} \gamma_{\rm o}^{t-m}
		}
		,
		& \!\!\!\! \text{if}\ t > 0,
		\\
		0,
		& \!\!\!\! \text{o.w.},
	\end{cases}
\end{align}
where $\beta_{\rm threshold}$ is the threshold for the LSF to be classified in a good condition, $0 < \gamma_{\rm o} \le 1$ is a discount factor for the channel classification history, and $\mathbbm{1}\{\cdot\}$ is the indicator function where $\mathbbm{1}\{A\} = 1$, if condition $A$ is valid, and equals $0$, otherwise. The term $\gamma_{\rm o}^{t-m}, m \le t$ found in the numerator of~\eqref{eq:zeta_history} prioritizes newer history.

The intuition from $\zeta_{bu}^{(t)}$ in both~\eqref{eq:zeta_direction} and~\eqref{eq:zeta_history} is to provide information about the possible status of the channels in the future for the DRL algorithm. This is very useful when the aim is to minimize the number of HOs or accumulate them while maintaining a good rate performance.

\subsection{Reward Function}
To formulate the reward function, it is necessary to account for HOs through a penalty term. We propose to use the following reward function.
\begin{align}\label{eq:reward_HOPenalty}
%
	\bar{R}\left({\bf o}^{(t)}, {\bf a}^{(t)}\right) = 
	\alpha^{(t)}
	R_u^{(\rm ind)}\left({\bf o}^{(t)}, {\bf a}^{(t)}\right)
\end{align}
where 
$R_u^{(\rm ind)}\left({\bf o}^{(t)}, {\bf a}^{(t)}\right)$ is the achievable rate that is an signal-to-noise ratio (SNR) based version of~\eqref{eq:rate_LB} (and is defined next), and $\alpha^{(t)}$ is the fractional time remaining for data transmission after we deduct the weighted time penalty for the HOs; this penalty parameter is defined as
\begin{align}
	\alpha^{(t)}
	=
	\frac{N_{\rm c} \tau_{\rm c} - \tau_{\rm HO, total}^{(t)}}{N_{\rm c} \tau_{\rm c}}
\end{align}
where $\tau_{\rm c}$ is the length of a communications cycle, $N_{\rm c}$ is the number of communication cycles contained in a single decision step, and $\tau_{\rm HO, total}^{(t)}$ is the \textit{weighted} overhead from executing the HOs at the beginning of decision step $t$. We define $\tau_{\rm HO, total}^{(t)}$ through a nonlinear function 
using
\begin{align}
	\tau_{\rm HO, total}^{(t)}
	&=
	\min \left(
	\mathbbm{1} \left\{N^{(t)} > 0 \right\} \tau_0 +  N^{(t)} \tau_{\rm HO}
	,\
	N_{\rm c} \tau_{\rm c}
	\right)
	\label{eq:HO_overhead}
	\\
	N^{(t)} &=
	\| [{\bf a}^{(t)} - {\bf a}^{(t-1)}]^+ \|^2
\end{align}
where $N^{(t)}$ gives number of added/removed APs at $t$, the term $\tau_0$ is a basic overhead for initiating HOs; it represents the weighted time spent doing radio resource control (RRC) reconfiguration, handshakes, reporting, etc. The term $\tau_{\rm HO}^{(t)}$ is the weighted overhead resulting from each HO, and $N^{(t)}$ is the number of HOs per decision step~$t$. 
The minimum operator found in~\eqref{eq:HO_overhead} simply indicates that the duration between any two decision steps $t-1$ and $t$ should be chosen to be more than the maximum duration of any HOs decision. The intuition from the nonlinear function in~\eqref{eq:HO_overhead} is to both allow for a flexible time overhead that is a function of the number of HOs and to use a penalty for initiating HOs.


The achievable data rate found in the reward function in~\eqref{eq:reward_HOPenalty} is based on~\eqref{eq:rate_LB}, and it is defined~as
\begin{align}\label{eq:reward}
	&R_u^{(\rm ind)}\left({\bf o}^{(t)}, {\bf a}^{(t)}\right) =
	\frac{1}{\tau_{\rm c}}
	\nonumber\\[-5pt]
	&
	\times 
	\sum_{n=n_{\rm est}}^{\tau_{\rm c}}
	\resizebox{0.85\columnwidth}{!}
	{$\displaystyle
	\log \Bigg( 1 + \frac{ \xi_{1,{\rm S}, u}\left[n, {\bf o}^{(t)}, {\bf a}^{(t)}\right]
	}{\displaystyle
		\xi_{2,3,{\rm S}, u}\left[n, {\bf o}^{(t)}, {\bf a}^{(t)}\right]	
		+ \hat{I}_{\rm term} \left[{\bf o}^{(t)}, {\bf a}^{(t)}\right]
		+ \sigma_{\rm z}^2 } \Bigg)
	$} 
\end{align}
where
{\allowdisplaybreaks
\begin{align}
	\xi_{1,{\rm S}, u}\left[n, {\bf o}^{(t)}, {\bf a}^{(t)}\right] 
	& =
	M^2
	\rho_{u}^2[n - n_{\rm est}]
	\Big|
	\sum_{b \in \mathcal{B}}
	a_{bu}^{(t)}
	\sqrt{\eta_{bu}^{(t)}} 
	\psi_{{\rm S}, bu}^{(t)}
	\Big|^2
	\label{eq:Reward_DS}
	\\[-8pt]
	\xi_{2,3,{\rm S}, u}\left[n, {\bf o}^{(t)}, {\bf a}^{(t)}\right]
	& =
	M^2
	\sum_{b \in \mathcal{B}}
	a_{bu}^{(t)}
	\eta_{bu}^{(t)} \beta_{bu}^{(t)} 
	\psi_{{\rm S}, bu}^{(t)}
	\label{eq:Reward_BU_CA}
	\\[-3pt]
	\hat{I}_{\rm term} \left[{\bf o}^{(t)}, {\bf a}^{(t)}\right]
	&
	= 
	p^{({\rm d})} 
	\sum_{b' \in \mathcal{B}}
	\left(
	1 - a_{b'u}^{(t)}
	\right)
	\beta_{b'u}^{(t)}
	\label{eq:emulatedInterf}
	%
\end{align}
}
%
%
%
%
%
$\!\!$with the variance of estimated channel is written as
\begin{align}
	\psi_{{\rm S}, bu}^{(t)}
	=
	\frac{\rho_{u}^2[ n_{\rm est} - i ] p^{({\rm u})} \big(\beta_{bu}^{(t)}\big)^2 }{ \sigma_{\rm z}^2 },
	\quad \text{for}\ u \in \mathcal{U}_i
	.
\end{align}
and the transmission power normalization $\eta_{bu}^{(t)}$ for user $u$ is
\begin{align}\label{eq:eta}
	\eta_{bu}^{(t)}
	=
	\frac{p^{({\rm d})}}
	{
		M ( |\mathcal{E}_b| + 1) \psi_{{\rm S}, bu}^{(t)}
	}
	.
\end{align}
where $p^{({\rm d})}$ is the downlink power budget of each AP. The expression in~\eqref{eq:emulatedInterf} provides an estimate of the interference experienced by the user from the non-served APs. This approximation enables the implementation of our HO solution on an individual basis for each user. Consequently, the function in~\eqref{eq:reward}, used in the reward in~\eqref{eq:reward_HOPenalty}, depends on the LSF and is predominantly determined by the SNR term while also factoring in the interference from non-serving APs. Additionally, the expression in~\eqref{eq:reward} is averaged over the small-scale fading because HO decisions cannot be based on the fast changing and highly random \textit{instantaneous} channels.

Using long-term statistics for taking HO decisions is more reliable than instantaneous channel realizations which fluctuate rapidly and could lead to frequent, unnecessary, HOs that disrupt the user's experience. Long-term statistics allow for strategic HOs decisions and help filter out momentary disruptions in the channels.

\section{Handoff Policy through DRL}\label{sec:DRL}
In this section, we construct a DRL framework that manages HOs as the user moves so that the reward is maximized. Our DRL framework will produce a trained DNN that determines the connections decisions of the user, hence the DNN will be the HO policy. 
In contrast to a tabular policy, a DNN can handle large problem sizes and complex environments.

\subsection{Continuous Output of DNN}\label{subsec:ContinousA}
Using DRL, we train a DNN to be used as the HO policy for the user. The output of the DNN characterizes a continuous vector $\widetilde{\bf a}^{(t)} \in \mathbb{R}^{B \times 1}$, which is used to determine the connection decisions ${\bf a}^{(t)}$ through the following mapping. First, the indices of the maximum $B_{\rm con}$ in $\widetilde{\bf a}^{(t)}$ are defined as $\mathcal{I}_{B_{\rm con}}$, where as noted earlier, $B_{\rm con}$ is the number of APs that will serve the user. Then, $a_b^{(t)} = 1$ if $b \in \mathcal{I}_{B_{\rm con}}$; else $a_b^{(t)} = 0$.

Using a vector of continuous rather than a discrete action space provides a scalable DRL framework and facilitates the training of the DNN. For example, for a network of $30$ APs, where the user needs to connect to $B_{\rm con} = 5$ APs, we need about $142506$ neurons on the output layer for a discrete space, while we need only $30$ neurons (or possibly double that depending on the DRL algorithm used) for a vector of continuous space. In the results section, we illustrate the benefits of using a continuous space.

\begin{figure}[t]
	\centering
	\includegraphics[width=1\columnwidth]{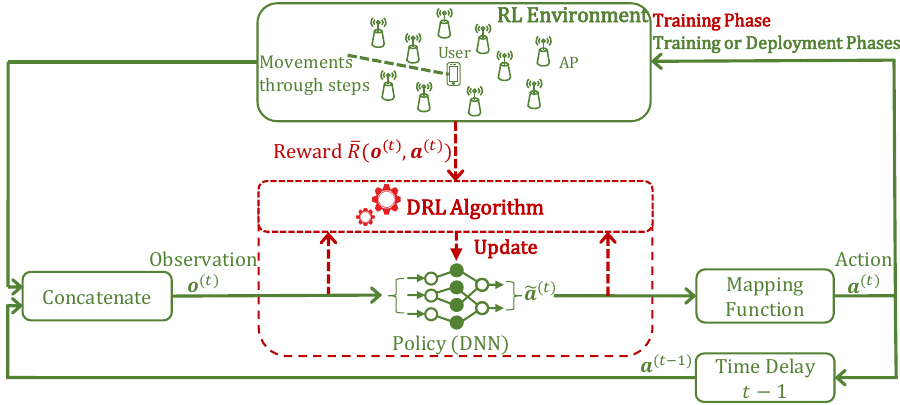}
	\vspace{-1.5em}
	\caption{DRL framework with a vector of continuous output $\widetilde{\bf a}^{(t)}$ for the DNN converted to the connection decisions ${\bf a}^{(t)}$.}
	\label{fig:mob_NNoutput_conversion}
	\vspace{-1.5em}
\end{figure}

Fig.~\ref{fig:mob_NNoutput_conversion} provides an overview of our DRL framework. 
The figure illustrates the construction of the observation ${\bf o}^{(t)}$ that serves as the input to the DNN, as well as the update process provided by the DRL algorithm to optimize the DNN parameters and develop an effective policy. We note that the time delay (bottom of Fig.~\ref{fig:mob_NNoutput_conversion}) implies that the previous connection decision form the input to the DNN.

\subsection{Soft Actor-Critic}
To train our DNN, we use the SAC algorithm with automatic tuning for the temperature parameter (discussed later) as proposed in~\cite{haarnoja2018soft}. The SAC algorithm is an off-policy learning technique, which means it reuses past experience through a replay buffer $\mathcal{D}$ to perform the training. This buffer saves past experiences from the RL environment, and it contains the following entries: [state ${\bf o}^{(t)}$, action ${\bf A}^{(t)}$, reward $\bar{R}({\bf o}^{(t)}, {\bf a}^{(t)})$, next-state ${\bf o}^{(t+1)}$, episode termination indicator]. Then, to train the DNNs, a batch (set of samples) is extracted through random sampling without repetition.

Compared to tabular policies, the SAC algorithm approximates a Q-function and the policy to avoid high computational complexities when dealing with large continuous domains. The Q-function indicates how good it is to perform a specific action in a specific state while following some policy. Moreover, the SAC algorithm jointly performs policy evaluation and policy improvement iteratively instead of running either of these operations till convergence. This simultaneous optimization is achieved through the use of two DNNs referred to as the Actor and the Critic.

\subsubsection{Objective}		
The SAC algorithm represents the policy as a parametric probability distribution $\pi_{\bm \Upsilon}({\bf a} | {\bf o}^{(t)}) = \mathbb{P}\{{\bf a}|{\bf o}^{(t)};{\bm \Upsilon}\}$ that stochastically selects action ${\bf a}$ given state ${\bf o}^{(t)}$ according to parameters ${\bm \Upsilon}$ (so we say the policy is parameterized by ${\bm \Upsilon}$). Specifically, the output of the policy is a vector of mean ${\bm \mu}_{\bm \Upsilon}( {\bf o}^{(t)} ) \in \mathbb{R}^{B \times 1}$ and log standard deviation $\log({\bm \sigma}_{\bm \Upsilon}({\bf o}^{(t)}) ) \in \mathbb{R}^{B \times 1}$ which together characterize a vector of normal distributions used to determine $\widetilde{\bf a}^{(t)}  \in \mathbb{R}^{B \times 1}$ and hence our HO decisions ${\bf a}^{(t)}$. Using the discussion in Section~\ref{sec:ActionSpace}, we define~${\bf a}^{(t)} \in \mathbb{B}^{B \times 1}$ using $\{ a_b^{(t)} = \mathbbm{1}\{ b \in \arg {\rm maxK}(\widetilde{\bf a}^{(t)}; B_{\rm con}) \} \}$, where $\arg {\rm maxK}({\bf x}; B_{\rm con})$ returns the $B_{\rm con}$ indices of the continuous vector ${\bf x}$ having the maximum values, and $\mathbbm{1}\{X\}$ is the indicator function which returns $1$ if the condition $X$ is true and $0$ otherwise. To simplify the discussion, we will use ${\bf a}^{(t)}$ as our output because it is the one directly used by the reward.

An optimal policy is a policy that maximizes the entropy objective function defined as
\begin{align}\label{eq:entropy_obj}
	J_{\rm O}(\pi_{\bm \Upsilon})
	=
	\resizebox{0.82\columnwidth}{!}
	{$\displaystyle
	\sum_{t = 0}^{T} \mathbb{E}_{({\bf o}^{(t)}, {\bf a}^{(t)}) \sim \rho_{\pi_{\bm \Upsilon}}}
	\left\{
	\bar{R}({\bf o}^{(t)}, {\bf a}^{(t)}) + \alpha_{\rm T} \mathcal{H}_{\pi_{\bm \Upsilon}} (\cdot|{\bf o}^{(t)}) \right\}
	$}
\end{align}
where ${\bf o}^{(t)}$ is the observation defined in~\eqref{eq:observation_state} and it represents the state, $\bar{R}(\cdot, \cdot)$ is our reward function defined in~\eqref{eq:reward_HOPenalty}, and $T$ is the time horizon. The term $\mathcal{H}_{\pi_{\bm \Upsilon}} (\cdot|{\bf o}^{(t)}) = - \mathbb{E}_{{\bf a}^{(t)}}[ \log \pi_{\bm \Upsilon}({\bf a}^{(t)}|{\bf o}^{(t)}) ] = - \sum_{{\bf a}^{(t)}} \pi_{\bm \Upsilon}({\bf a}^{(t)} | {\bf o}^{(t)}) \log \pi_{\bm \Upsilon}({\bf a}^{(t)} | {\bf o}^{(t)})$ is the entropy function of the stochastically selected action ${\bf a}^{(t)}$ using the probability distribution $\pi_{\bm \Upsilon}(\cdot | {\bf o}^{(t)})$ evaluated at state ${\bf o}^{(t)}$. The term $\alpha_{\rm T} \ge 0$ is called the temperature parameter and it determines the relative importance of the entropy with respect to the reward, and is used to control the stochasticity of the optimal policy. The term $\rho_{\pi_{\bm \Upsilon}}$ is the joint distribution of the states and actions, and is induced by the policy being trained. In practice, we will use the replay buffer $\mathcal{D}$ to draw these samples $({\bf o}^{(t)}, {\bf a}^{(t)})$ and empirically estimate the distribution $\rho_{\pi_{\bm \Upsilon}}$. As discussed in~\cite{haarnoja2018soft}, the objective function in~\eqref{eq:entropy_obj} can incorporate a discount factor $\gamma$. It is essential to include this discount factor when the time horizon $T$ tends to infinity.

The entropy term in~\eqref{eq:entropy_obj} encourages the policy to have a level of randomness which increases the exploration of actions and prevents premature convergence to suboptimal solutions. The temperature parameter $\alpha_{\rm T}$ scales the entropy term with respect to the reward; this scaling can be static or using the automatic tuning method~\cite{haarnoja2018soft}.

To obtain a policy that maximizes the objective function in~\eqref{eq:entropy_obj}, SAC employs a parameterized state value function $V_{\bm \Lambda}({\bf o}^{(t)})$, a parameterized \textit{soft} Q-function $Q_{\bm \Omega}({\bf o}^{(t)}, {\bf a}^{(t)})$ and a parameterized policy $\pi_{\bm \Upsilon} ({\bf a}^{(t)} | {\bf o}^{(t)})$ with respective parameters ${\bm \Lambda}$, ${\bm \Omega}$ and ${\bm \Upsilon}$. A soft Q-function is obtained starting from any Q-function and repeatedly applying the modified Bellman backup operator~\cite{haarnoja2018soft}.

\subsubsection{Actor and Critic}
The SAC algorithm represents both the policy and the soft Q-function as DNNs called the Actor and the Critic, respectively. The Actor represents our HO policy $\pi_{\bm \Upsilon} ({\bf a}^{(t)} | {\bf o}^{(t)})$ parameterized by a DNN with parameters ${\bm \Upsilon}$. For the Critic, we have two types of DNNs denoted as the Critic-local and Critic-target. The Critic-local represents two soft Q-functions $Q_{\bm \Omega_1}({\bf o}^{(t)}, {\bf a}^{(t)})$ and $Q_{\bm \Omega_2}({\bf o}^{(t)}, {\bf a}^{(t)})$ parameterized by two DNNs with parameters $\bm \Omega_1$ and $\bm \Omega_2$, respectively. Similarly, the Critic-target represents two target soft Q-functions $Q_{{\bf \bar{\bm \Omega}}_1}({\bf o}^{(t+1)}, {\bf a}^{(t+1)})$ and $Q_{{\bf \bar{\bm \Omega}}_2}({\bf o}^{(t+1)}, {\bf a}^{(t+1)})$ parameterized by two target DNNs with parameters ${\bf \bar{\bm \Omega}}_1$ and ${\bf \bar{\bm \Omega}}_1$, respectively.


The two Critic DNNs are trained independently. Using two soft Q-functions, and taking the minimum value between the pair, mitigates the positive bias (overestimation) in the policy improvement step and hence prevents the degradation of performance~\cite{fujimoto2018addressing}. Moreover, it significantly speeds up the training, especially on hard-to-learn tasks. After an update for these DNNs, the minimum soft Q-function is used for the value of the gradients as discussed next.

\subsubsection{Critic DNNs}
The parameters $\{\bm \Omega_i : i = 1, 2\}$ of the soft Q-function can be trained to minimize the mean squared error 
(loss) written as
\begin{align}\label{eq:critic_loss}
	J_Q(\bm \Omega_i)
	&=
	\resizebox{0.82\columnwidth}{!}
	{$\displaystyle
	\mathbb{E}_{({\bf o}^{(t)}, {\bf a}^{(t)}) \sim \mathcal{D}  }
	\left[
	\left(
	Q_{\bm \Omega_i}({\bf o}^{(t)}, {\bf a}^{(t)})
	-
	\hat{Q}({\bf o}^{(t)}, {\bf a}^{(t)})
	\right)^2
	\right]
	$}
	\nonumber \\
	&
	\quad
	,
	i = 1, 2
\end{align}

The term $\hat{Q}({\bf o}^{(t)}, {\bf a}^{(t)})$ is defined as
\begin{align}
	\hat{Q}({\bf o}^{(t)}, {\bf a}^{(t)})
	=
	\bar{R}({\bf o}^{(t)}, {\bf a}^{(t)})
	+
	\gamma \mathbb{E}_{ {\bf o}^{(t+1)} 
	\sim p_{\rm o} }
	[V_{\bm \Lambda}({\bf o}^{(t+1)})],
\end{align}
where 
${\bf o}^{(t+1)}$ is the next-state,
$p_{\rm o}$ is the transition probability from the current-state to the next-state, and $V_{\bm \Lambda}({\bf o}^{(t+1)})$ is the state value function that can be parameterized through a soft Q-function target parameters using 
\begin{align}\label{eq:valueFunction}
	V_{\bm \Lambda}({\bf o}^{(t+1)})
	&=
	Q_{{\bf \bar{\bm \Omega}}, \min}({\bf o}^{(t+1)}, {\bf a}^{(t+1)})
	\nonumber \\
	&\quad
	-
	\alpha_{\rm T} \log \pi_{\bm \Upsilon} ({\bf a}^{(t+1)} | {\bf o}^{(t+1)} ),
\end{align}
where ${\bf a}^{(t+1)}$ is obtained using the \textit{reparameterization trick}~\cite{haarnoja2018soft2} in which a sample from $\pi_{\bm \Upsilon}(\cdot|{\bf o}^{(t+1)})$ is drawn by computing a deterministic function of the next-state ${\bf o}^{(t+1)}$, policy parameters ${\bm \Upsilon}$, and independent noise $\epsilon$ as
\begin{align}
{\bf a}^{(t+1)} = \mathtt{tanh}({\bm \mu}_{\bm \Upsilon}({\bf o}^{(t+1)}) + {\bm \sigma}_{\bm \Upsilon}({\bf o}^{(t+1)}) \odot \epsilon ) ,\  \epsilon \sim \mathcal{N}(0, I)
\end{align}
and $Q_{{\bf \bar{\bm \Omega}}, \min}(\cdot, \cdot)$ is the minimum of two soft target Q-functions. For notation simplicity, we define
\begin{align}\label{eq:minQ}
	Q_{{\bf X}, \min}({\bf o}^{(t)}, {\bf a}^{(t)})
	&=
	\min(
	Q_{{\bf X}_1}({\bf o}^{(t)}, {\bf a}^{(t)}),
	Q_{{\bf X}_2}({\bf o}^{(t)}, {\bf a}^{(t)}))
	\nonumber \\
	&\quad \quad
	,
	{\bf X} \in \{{\bf \bar{\bm \Omega}}, \bm \Omega \}
\end{align}

The state value function can be related to the soft Q-function and the policy through~\eqref{eq:valueFunction}, however, in older versions of SAC, a separate DNN is used as a function approximator.

As noted earlier, the terms $\{Q_{{\bf \bar{\bm \Omega}}_i}({\bf o}^{(t+1)}, {\bf a}^{(t+1)}) : i=\{1,2\}\}$ defined through $Q_{{\bf \bar{\bm \Omega}}, \min}({\bf o}^{(t+1)}, {\bf a}^{(t+1)})$, where $(\cdot)_{x, \min}$ is defined in~\eqref{eq:minQ}, are two target soft Q-functions with parameters $\{ {\bf \bar{\bm \Omega}} : i=\{1,2\} \}$. These functions are represented through two Critic-target DNNs and updated using an exponentially moving average of the soft Q-function weights (soft update), i.e., ${\bf \bar{\bm \Omega}}_i \leftarrow T_{\rm sm} {\bm \Omega}_i - (1 - T_{\rm sm}) {\bf \bar{\bm \Omega}}_i $ for $i\in \{1, 2\}$, where $T_{\rm sm} << 1$ is a smoothing coefficient.


\subsubsection{Actor DNN}

The actor DNN represents our policy that determines the HO decisions (actions). The policy is modeled as a vector of normal distributions with mean ${\bm \mu}_{\bm \Upsilon}( {\bf o}^{(t)} )$ and log standard deviation $\log({\bm \sigma}_{\bm \Upsilon}({\bf o}^{(t)}) )$ given by the Actor DNN. The policy parameters ${\bm \Upsilon}$ can be updated by minimizing the expected relative entropy (Kullback-Leibler (KL) divergence) ${\rm D}_{\rm KL} \left(p||q\right)$, between the current policy and the desired policy, multiplied by the temperature term $\alpha_{\rm T}$, as
\begin{align}\label{eq:modified_KL_divergence}
	J_\pi ({\bm \Upsilon})
	&=
	\mathbb{E}_{{\bf o}^{(t)} \sim 
	\mathcal{D}
	}
	\left[
	\mathbb{E}_{{\bf a}^{(t)} \sim \pi_{\bm \Upsilon}}
	\left[
	\alpha_{\rm T} \log \pi_{\bm \Upsilon} ( {\bf a}^{(t)} | {\bf o}^{(t)} )
	\right. \right.
	\nonumber \\
	&\quad \quad \quad \quad \quad \quad \quad \quad \quad
	\left. \left.
	-
	Q_{\bm \Omega, \min} ({\bf o}^{(t)}, {\bf a}^{(t)})
	\right]
	\right],
\end{align}
where ${\bf a}^{(t)} \sim \pi_{\bm \Upsilon}$ are sampled using the current policy instead of the replay buffer $\mathcal{D}$.

The policy $\pi_{\bm \Upsilon} ( {\bf a}^{(t)} | {\bf o}^{(t)} )$ can be reparameterized using a DNN transformation written as ${\bf a}^{(t)} = f_{\bm \Upsilon}(\epsilon^{(t)}; {\bf o}^{(t)})$, where this operation is known as the \textit{reparameterization trick}~\cite{haarnoja2018soft2}. 
This implies that ${\bm \Upsilon}$ can be updated by minimizing the expected relative entropy written as follows
\begin{align}\label{eq:modified_KL_divergence2}
	J_\pi ({\bm \Upsilon})
	&=
	\mathbb{E}_{{\bf o}^{(t)} \sim 
	 \mathcal{D}
	, \epsilon^{(t)} \sim \mathcal{N}(0, I)} \left[
	\alpha_{\rm T} \log \pi_{\bm \Upsilon} ( f_{\bm \Upsilon}(\epsilon^{(t)}; {\bf o}^{(t)}) | {\bf o}^{(t)} )
	\right.
	\nonumber \\
	& \quad \quad \quad \quad \quad \quad
	\left.
	-
	Q_{\bm \Omega, \min} ({\bf o}^{(t)}, f_{\bm \Upsilon}(\epsilon^{(t)}; {\bf o}^{(t)}))
	\right],
\end{align}
where $\epsilon^{(t)}$ is an input noise vector sampled from some fixed distribution such as the spherical Gaussian distribution $\mathcal{N}$.

\begin{figure}[t]
	\centering
	\includegraphics[width=1\linewidth]{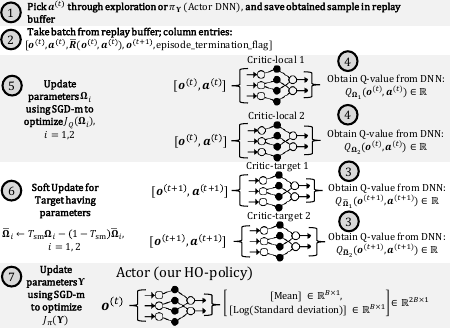}
	\vspace{-1em}
	\caption{SAC operations used to obtain our HO policy represented by actor DNN. Irrelevant details are removed for 
		clarity.}
	\label{fig:actor_critic}
	\vspace{-1em}
\end{figure}

\subsubsection{Training}
Stochastic gradient descent with momentum (SGD-m), e.g., the Adaptive Moment Estimation (ADAM), is used to optimize the parameters $\{\bm \Omega_i: i=\{1,2\}\}$ for the Critic-local DNNs and $\{{\bm \Upsilon}: i=\{1,2\} \}$ for the Actor DNN, based on the loss functions in~\eqref{eq:critic_loss} and~\eqref{eq:modified_KL_divergence2}, respectively. The parameters $\{{\bf \bar{\bm \Omega}}_i: i=\{1,2\} \}$ of the Critic-target DNNs are updated as ${\bf \bar{\bm \Omega}}_i \leftarrow T_{\rm sm} {\bm \Omega}_i - (1 - T_{\rm sm}) {\bf \bar{\bm \Omega}}_i $ with soft update parameter $T_{\rm sm}$ as noted earlier. One or several gradient steps are usually performed per each RL environment decision step~$t$. The method also alternates between collecting experiences from the environment with the current policy and updating the function approximators using both the SGD-m and batches sampled from the replay buffer.

It is worth noting that, in our implementation, the state value function is updated implicitly through~\eqref{eq:valueFunction}, rather than using an additional DNN as a function approximator. We also use the automatic tuning method for the temperature parameter $\alpha_{\rm T}$ that is performed through a separate DNN as described in~\cite[Section  5]{haarnoja2018soft}. Fig.~\ref{fig:actor_critic} summarizes the training steps using the SAC algorithm for our HO management framework. The main steps can be decomposed into \textbf{1)} calculating the Critic loss and updating the Critic parameters $\big($Steps~$\raisebox{.5pt}{\textcircled{\raisebox{-.9pt} {3}}}-\raisebox{.5pt}{\textcircled{\raisebox{-.9pt} {6}}} \big)$, and \textbf{2)} calculating the Actor loss and updating the Actor parameters $\big($Step~$\raisebox{.5pt}{\textcircled{\raisebox{-.9pt} {7}}} \big)$. Once the DNN framework is trained, the Actor DNN is used to manage the HO decisions of the user, where the output of the Actor DNN that corresponds to the mean (Fig.~\ref{fig:actor_critic}) is used as the vector of continuous space~$\widetilde{\bf a}^{(t)}$ introduced in Section~\ref{subsec:ContinousA}.

\subsection{Deployment and Location of DRL framework}\label{section:ActorCriticLoc}
\vspace{-0.2em}
The DRL framework is flexibile in terms of its implementation location. This is due to the fact that it can be trained offline. A candidate deployment scenario is to deploy the DRL framework on the network side and perform the training offline. After that there is freedom to either keep the obtained HO policy (Actor DNN) on the network side to determine the HO decisions for each user individually, or to distribute copies of the HO policy (Actor DNN) to the users to make them perform HO decisions directly. Both options should be feasible because once the DNN is trained we can take the optimized HO decisions in a very short time with small computational power. We will show our measurements for the delay to obtain HO decisions in the Results section.

\begin{table*}[t]
	\scriptsize
	\centering
	\caption{Simulation parameters for the network.}
	\vspace{-0.7em}
	\begin{tabular}{|p{0.29\linewidth}|p{0.14\linewidth}||p{0.29\linewidth}|p{0.14\linewidth}|}
		\hline
		\hline
		\multicolumn{1}{|l|}{ \textit{\textbf{Parameter}}} & \multicolumn{1}{l||}{\textit{\textbf{Value}}}&\multicolumn{1}{l|}{ \textit{\textbf{Parameter}}} & \multicolumn{1}{l|}{\textit{\textbf{Value}}}\\
		\hline
		Number of APs: $B$ & $27$ APs
		&
		Duration between $t$ and $t+1$: $\bar{\Delta}$ & $5~{\rm s}$
		\\
		Number of antennas per AP: $M$ & $8$ antennas
		&
		Path loss terms: $\alpha_{\rm pl}$, $d_0$, $d_{\rm h}$.
		& $3.8$, $1.1~{\rm m}$, $13.5~{\rm m}$
		\\
		Number of serving APs: $B_{\rm con}$ & $5$ APs
		&
		Speed of light: $c_0$ & $3\times 10^8~{\rm m/s}$
		\\
		%
		Downlink transmit power: $p^{({\rm d})}$ & $30~{\rm dBm}$ 
		&
		Sampling period $T_{\rm s}$ & $66.7~\mu{\rm s}$
		\\
		Uplink transmit power $p^{({\rm u})}$ & $20~{\rm dBm}$
		&
		Noise spectral density 
		& $-174~{\rm dBm/ Hz}$
		\\
		Length of communication cycle $\tau_{\rm c}$ &
		$200$ channel uses
		&
		Noise figure 
		& $8~{\rm dB}$
		\\
		Length of pilot training phase $\tau_{\rm p}$ & $16$ channel uses
		&
		Std. deviation of shadowing: $\sigma_\mathrm{sh|dB}$& $6~{\rm dB}$
		\\
		Bandwidth & $2~{\rm MHz}$
		&
		Correlated shadowing: $d_\mathrm{decorr}$, $\iota$ & $100~{\rm m}$, $0.5$
		\\
		Carrier frequency: $f_{\rm c}$ & $1.8~{\rm GHz}$
		&
		Threshold for good channel: $\beta_{\rm threshold}$ & ${\rm PL}(300~{\rm m})$
		\\
		User's speed: $\mathtt{v}_u$ & $10~{\rm m/s}$
		&
		Discount factor for channel history: $\gamma_{\rm o}$ & 0.8
		\\
		Number of users served by AP: $\mathcal{E}_b\in \mathbb{Z}$ & $0 \le \mathcal{E}_b \le 5$, $\mathcal{E}_b$
		&
		Avg. number of users served by AP: $\mu_{|\mathcal{E}|}$ & $3$ users
		\\
		\hline
		\hline
	\end{tabular}
	\label{table:sim_parameters}   
	\vspace{-0.5em}
\end{table*}

\begin{table*}[t]
	\tiny
	\centering
	\caption{Hyperparameters used by SAC.}
	\vspace{-1em}
	\begin{tabular}{|p{0.22\linewidth}|p{0.05\linewidth}||p{0.22\linewidth}|p{0.05\linewidth}|			|p{0.22\linewidth}|p{0.05\linewidth}|}
		\hline
		\hline
		\multicolumn{1}{|l|}{ \textit{\textbf{Hyperparameter}}} & \multicolumn{1}{l|}{ \textit{\textbf{Value}}}
		&
		\multicolumn{1}{l|}{ \textit{\textbf{Hyperparameter}}} & \multicolumn{1}{l|}{ \textit{\textbf{Value}}}
		&
		\multicolumn{1}{|l|}{ \textit{\textbf{Hyperparameter}}} & \multicolumn{1}{l|}{ \textit{\textbf{Value}}}
		\\
		\hline
		Number of hidden layers & 2
		&
		Hidden layers activation & Relu
		&
		Final layer activation & None 
		\\
		Number of neurons & [64, 64]
		&
		buffer size & 1000000
		&
		Optimizer & Adam
		\\
		Discount rate: $\gamma$ & 0.99
		&
		Batch size & 256
		&
		Min. decision steps before learning & 400
		\\
		Learning rate & 0.0001
		&
		Automatically tuning for temperature: $\alpha_{\rm T}$ & True
		&
		Initializer & Xavier
		\\
		Target smoothing coefficient: $T_{\rm sm}$ & 0.005
		&
		Gradient clipping norm & 5
		&
		&
		\\
		Replay buffer size (Critic-specific) & 1000000
		&
		Final layer activation (Actor-specific) & Tanh
		&
		&
		\\
		\hline
		\hline
	\end{tabular}
	\label{table:sim_parameters_NN_AC}   
	\vspace{-3em}
\end{table*}

\vspace{-0.6em}
\section{Partial Observability}\label{sec:POMDP}
\vspace{-0.2em}
When the state of the RL environment cannot be directly observed, the observation will represent incomplete or noisy information about the state. In this case, the system is partially observable. Formally, our RL environment is defined through a POMDP instead of an MDP.

Using POMDP makes our system more practical for implementation, because it does not have strict requirements, e.g., in our case it can be implemented without knowing the LSF to \textit{all} the APs in the network. Accordingly, our system can be more flexible in the sense it can maintain its functionality with limited channel state information.


The partial observability feature could be useful in many scenarios. One possible scenario is when the APs are being deactivated/activated to save energy consumption. In such a case, an up-to-date LSF statistics will not be available, and partial observability for the state of the RL environment needs to be applied. Another scenario is when the LSF statistics are noisy due to noisy observations.

We assume that when the user is not connected to a AP, then we do not have access to the exact information concerning this AP. In this case, we define our partially observable state through the observation vector ${\bf o}^{(t)}_{\rm PO} \in \mathbb{R}^{4 B \times 1}$ as:
\vspace{-0.5em}
\begin{align}\label{eq:observation_POMDP}
	{\bf o}^{(t)}_{\rm PO} &= \left[ {\rm S}\{\log(\widetilde{\beta}_{1u}^{(t)}) \ \dots \ \log(\widetilde{\beta}_{Bu}^{(t)} ) \} \ {\rm S}\{|\widetilde{\mathcal{E}}_1| \ \dots \ |\widetilde{\mathcal{E}}_B| \} 
	\right.
	\nonumber \\[-5pt]
	& \quad \quad
	\left.
	{\rm S}\{ a_{1u}^{(t-1)} \ \dots \ a_{Bu}^{(t-1)} \} 
	\ {\rm S}\{ \zeta_{1u}^{(t)} \ \dots \ \zeta_{Bu}^{(t)} \} \right]^T
\end{align}
where
\vspace{-1em}
\begin{align}
	\left( \widetilde{\beta}_{bu}^{(t)}, |\widetilde{\mathcal{E}}_b | \right)
	=
	\begin{cases}
		\left(\beta_{bu}^{(t)}, |\mathcal{E}_b | \right) , & \text{if}\ a_{bu}^{(t-1)} = 1
		\\
		\left( {\rm PL}(\bar{d}_{bu}^{(t)}) , \mu_{|\mathcal{E}|} \right), & \text{if}\ a_{bu}^{(t-1)} = 0
	\end{cases}
\end{align}
where $\mu_{|\mathcal{E}|}$ is the average number of users set to be served by a AP, and ${\rm PL}(\bar{d}_{bu}^{(t)})$ is the path loss between the user and AP $b$ which depends on the distance $\bar{d}_{bu}^{(t)}$. Using the path loss instead of the actual LSF reflects the absence of pertinent information, which could potentially impact the HO decisions.

As for the reward, we use the reward $\bar{R}\left({\bf o}^{(t)}, {\bf a}^{(t)}\right)$ in~\eqref{eq:reward_HOPenalty}, which depends on the actual state of the system defined in~\eqref{eq:observation_state}, rather than on the partial observable state defined in~\eqref{eq:observation_POMDP}.

\section{Numerical Results and Analysis}\label{sec:results}
We implement the RL environment that models the management of HOs using the OpenAI's $\mathtt{Gym/Gymnasium}$ library~\cite{brockman2016openai}. 
In this setup, an episode starts by generating a network where the user moves in discrete time steps with a velocity $\mathtt{v}_u$ and time resolution $\bar{\Delta}$. Each time step represents a decision step $t$ which includes connecting to $B_{\rm con}$ APs and obtaining new observation and reward. After the user moves for a distance of~$1~{\rm km}$, which corresponds to $20$~steps for a user's velocity of~$10~{\rm m/s}$, the episode terminates, i.e., the RL game ends, and another episode can start.

As the user moves, we perform network wraparound to eliminate the network's border effect and hence simulate a network with infinite area. When the location of a AP that \textit{is still serving the user} changes due to wraparound, we assume that a HO has occurred, which is logical because in a actual deployment the AP with a new location would function as a new AP. This emulation facilitates the creation of a network with infinite area and a realistic HO operation.

Our simulation consists of two parts:
\begin{enumerate}
\item In the first part, we train the actor DNN (Fig.~\ref{fig:actor_critic}) to serve as our HO policy. This training is performed over many episodes through the DRL framework implemented by the SAC algorithm. To ensure a efficient policy, we save the DNN whenever we achieve a new maximum rolling accumulated (sum) reward for an episode. We set the number of maximum episodes to be in the range of $1$~million episodes. 
The rolling accumulated reward is the average of the episode's total rewards received over the past few episodes, calculated using a window size which we configure to $100$~episodes, and it can be used to evaluate the training behavior of our DRL framework. Once the training is done, the last saved DNN 
is deployed to act as the HO policy for the user. 

\item In the second part, we are done from training and we evaluate the performance of our HO policy using Monte Carlo simulations. In this setup, the actor DNN is used to take the HO decisions for $2000$ new episodes. As indicated earlier, in each episode the user needs to travel $20$~steps; hence, the results correspond to $40000$ channel realizations with the existence of temporal events.
\end{enumerate}

\vspace{-1em}
\subsection{Spatial correlation of large-scale fading}\label{section:spatialCorreLSF}

We define the large-scale fading between AP $b$ and user $u$ as
$\beta_{bu}^{(t)} \triangleq {\rm PL}(\bar{d}_{bu}^{(t)}) \times 10^{\frac{\sigma_\mathrm{sh|dB} \bar{\kappa}_{bu}^{(t)}}{10}}$, 
where ${\rm PL}(\cdot)$ accounts for the effect of the path loss on the channel gain, and the second term accounts for the shadowing where $\sigma_\mathrm{sh|dB}$ is the standard deviation of the shadowing in dB scale and $\bar{\kappa}_{bu}^{(t)}$ is defined using the two-component shadowing model~\cite{CorrelatedShadowing104090}. Specifically, due to the presence of common obstacles (e.g., large buildings) in the network area before and after the user's movement, the shadowing term is correlated. Hence, we define the shadowing experienced by user at time $t$ as
\begin{align}\label{eq:shad_model}
	\bar{\kappa}_{bu}^{(t)} =
	\begin{cases}
		\sqrt{\iota} \kappa_{1,b} + \sqrt{1 - \iota} \kappa_{2}, & \text{if}\ t = 0\\ 
		\sqrt{\iota} \kappa_{1,b} + \sqrt{1 - \iota} \kappa_{2}^{(t)}, 
		& \text{if}\ t > 0
	\end{cases}	
\end{align}
where $\kappa_{1,b}, \kappa_{2}, \kappa_{2}^{(t)} \sim \mathcal{N}(0, 1)$, and $0 \le \iota \le 1$. The term $\kappa_{1,b}$ models the contributions of the obstacles in the vicinity of AP $b$ to the shadowing experienced by the user, while $\kappa_{2}$ and $\kappa_{2}^{(t)}$ model those in the vicinity of the user and affects the shadowing to all the APs. The covariance functions between $\kappa_{1,b}$ and $\kappa_{1,b'}$, and between $\kappa_{2,u}^{(t)}$ and $\kappa_{2,u'}^{(t-1)}$ are, respectively, given by
\begin{align}\label{eq:correl_shad}
	C_{\kappa_{1,b},\kappa_{1,b'}} &= \mathbb{E}\left\{\kappa_{1,b} \kappa_{1,b'} \right\} = 2^{-\frac{d_{1,b b'}}{ d_\mathrm{decorr}}}\ \text{and} 
	%
	\nonumber \\[-3pt]
	\  C_{ \kappa_{2,u}^{(t)}, \kappa_{2,u}^{(t-1)} } &= \mathbb{E}\left\{\kappa_{2,u}^{(t)} \kappa_{2,u}^{(t-1)} \right\} =
	2^{-\frac{\mathtt{v}_u \bar{\Delta} }{ d_\mathrm{decorr}}},\ t>0
\end{align}
The parameter $d_{1,b b'}$ is the distance between APs $b$ and $b'$, and $(\mathtt{v}_u \bar{\Delta})$ is the distance between the user at decision step $t$ and $t-1$. The decorrelation distance $d_\mathrm{decorr}$ has typical values between $20$ and $200$ ${\rm m}$, and it corresponds to the distance at which the correlation drops to $50\%$.


\begin{figure*}[t]
	\centering
	\begin{subfigure}[t]{0.32\textwidth}
		\centering
		\includegraphics[width=1\textwidth]{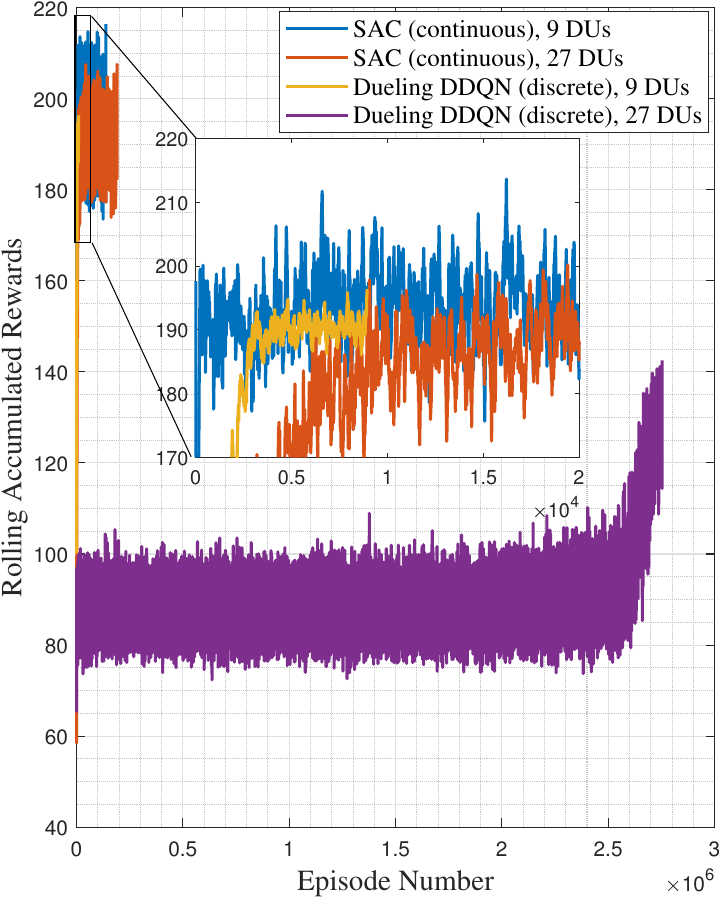}
		\vspace{-1.5em}
		\caption{Training.}
		\label{sfig:noPenalty_Training_compiled}
	\end{subfigure}
		$\ $
	\begin{subfigure}[t]{0.32\textwidth}
		\centering
		\includegraphics[width=1\textwidth]{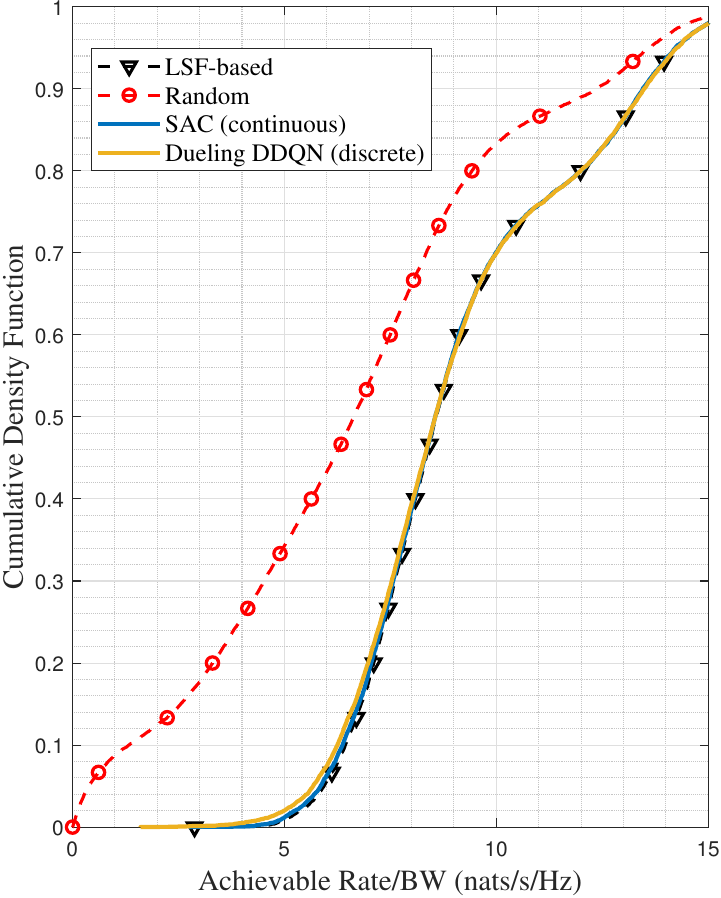}
		\vspace{-1.5em}
		\caption{Number of APs $B = 9$.}
		\label{sfig:noPenalty_CDF_9DUs}
	\end{subfigure}
			$\ $
	\begin{subfigure}[t]{0.32\textwidth}
		\centering
		\includegraphics[width=1\textwidth]{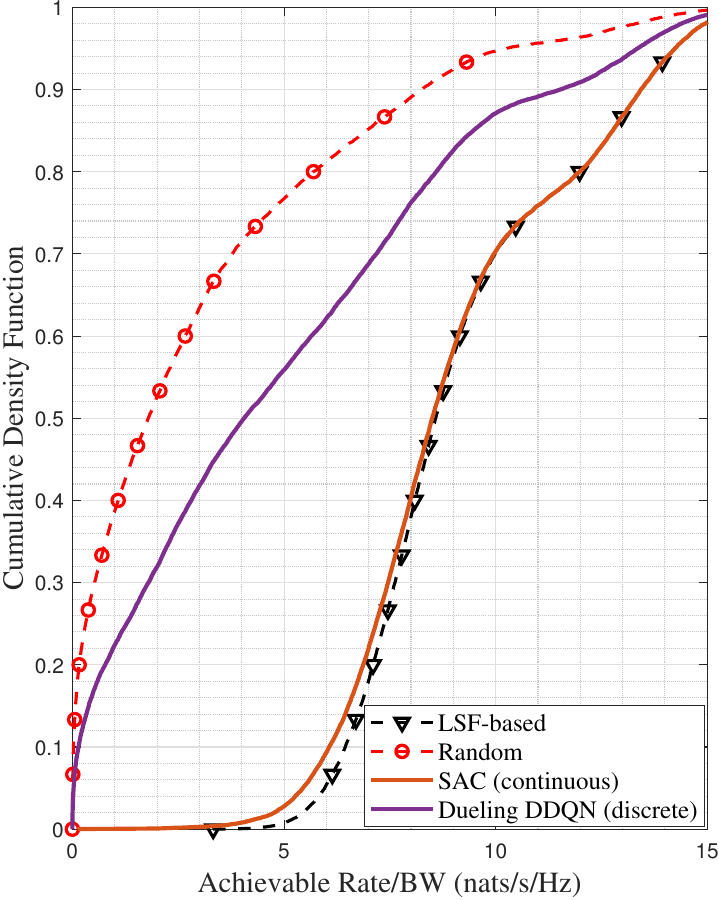}
		\vspace{-1.5em}
		\caption{Number of APs $B = 27$.}
		\label{sfig:noPenalty_CDF_27DUs}
	\end{subfigure}
	\vspace{-0.5em}
	\caption{Testing the hyperparameters for efficiency using: no HO penalty ($\alpha^{(t)} = 1$), $B_{\rm con} = 5$, equal number of served users, rolling window of size $100$ episodes.}
	\label{fig:noPenalty}
	\vspace{-1em}
\end{figure*}

Finally, we define the path loss using the COST231 Walfisch-Ikegami model~\cite{Walfisch14401} as
${\rm PL}(\bar{d}_{bu}^{(t)}) =
\bigg(\frac{\sqrt{(\bar{d}_{bu}^{(t)})^2 + d_{\rm h}^2}}
{d_0}
\bigg)^{-\alpha_{\rm pl} }$
%
%
%
%
%
%
where $\bar{d}_{bu}^{(t)}$ is the distance in the \emph{xy-plane} between AP $b$ and user $u$ at time index $t$, $\alpha_{\rm pl}$ is the path loss exponent, and $d_0$ is the reference distance. The term $d_{\rm h}$ is the fixed difference in the heights of the APs and the user; it is used to impose an exclusion region between the APs and the moving user.

The values of the parameters used for our simulations are presented in Table~\ref{table:sim_parameters} alongside the hyperparameters for DRL in Table~\ref{table:sim_parameters_NN_AC}. These values are used unless specified otherwise. In our simulations, the number of users assumed to be served by each AP, i.e., $|\mathcal{E}_b|, b \in \mathcal{B}$, is generated as uniform random variables between $0$ and $5$. 
In our single-user HO scheme, this number is used in the observation vector defined in~\eqref{eq:observation_state}, and it will affect the power allocation of the APs for the users, as shown in~\eqref{eq:eta}. By using this information, the DRL framework takes into account the load of the APs when deciding on the HO decisions. As future work, in a multi-user joint HO scheme, this information can prevent oveloading the APs with users while keeping other APs idle.

For the hyperparameters in Table~\ref{table:sim_parameters}, we choose the dimension of the DNNs as two hidden layers, each with 64 neurons. This size aligns well with the size of the observation (input layer) and the action (output layer) vectors. From our training results, this has demonstrated excellent performance given the complexity of our problem.

\subsection{Evaluation of DRL and the HO policy}\label{subsection:results_evaluation_DRL}
We compare our framework with the following:
\begin{itemize}
	\item \textbf{Random HO policy:} At each decision step $t$, the user connects randomly to $B_{\rm con}$ APs. This approach embodies an untrained HO policy, and by comparing it to our scheme we are able to measure the extent to which our framework has been trained.
	\item \textbf{LSF-based HO policy:} At each decision step $t$, the user connects to the $B_{\rm con}$ APs having the best LSF statistics. If HO decisions have no overhead, then this is the global optimum solution. When there is an overhead for HOs, then this is not the best HO policy. With partial observability, this policy cannot be applied.
	\item \textbf{DRL-based HO policy using discrete action space:} This approach uses our model but with a discrete action space. For the DRL algorithm, we use the dueling double DQN (dueling DDQN). The motivation to consider this approach is to contrast the scalability of our solution that uses a continuous action space.  
\end{itemize}


\begin{figure*}[t]
	\centering
	\begin{minipage}{.4\textwidth}
		\begin{subfigure}[t]{1\textwidth}
			\centering
			\includegraphics[width=1\textwidth]{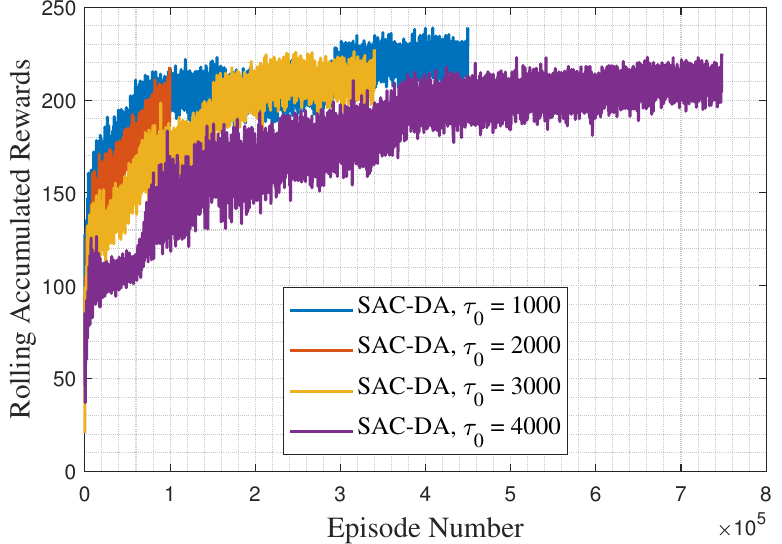}
			\vspace{-1.5em}
			\caption{Training.}
			\label{sfig:Training_DA_P_S}
		\end{subfigure}
		\\
		\begin{subfigure}[t]{1\textwidth}
			\centering
			\includegraphics[width=1\textwidth]{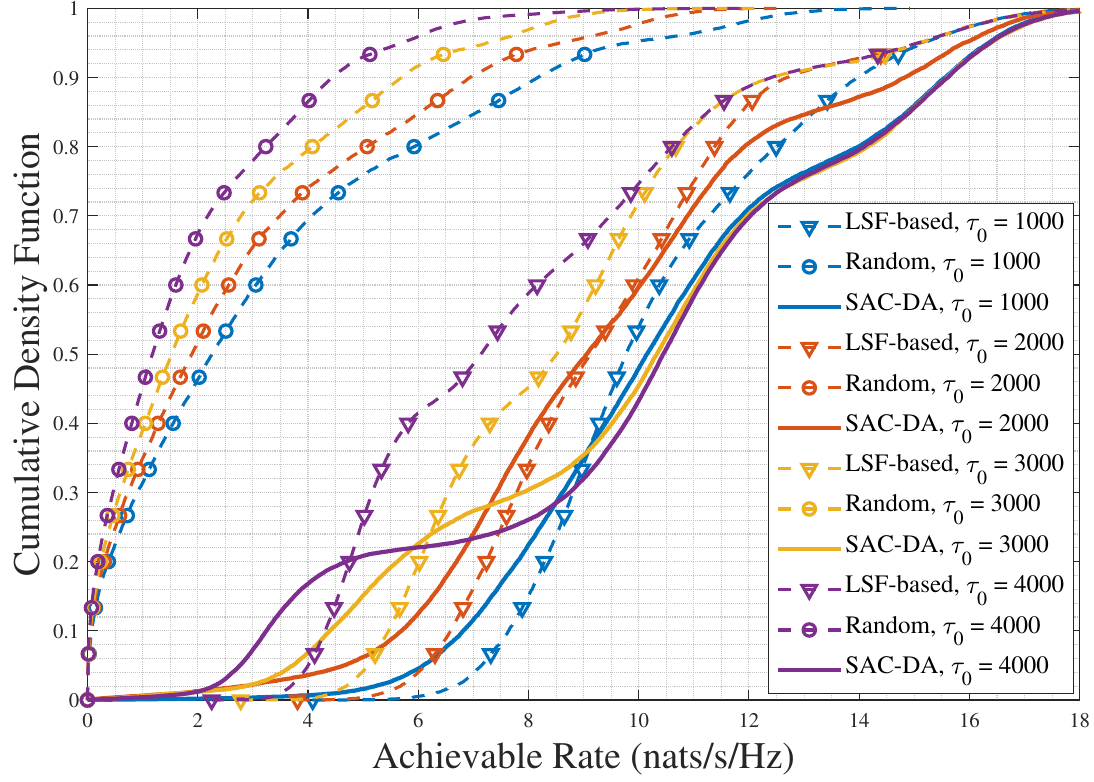}
			\vspace{-1.5em}
			\caption{CDF of achievable rate during the evaluation.}
			\label{sfig:CDF_DA_P_S}
		\end{subfigure}
		\\
		\begin{subfigure}[t]{1\textwidth}
			\centering
			\includegraphics[width=1\textwidth]{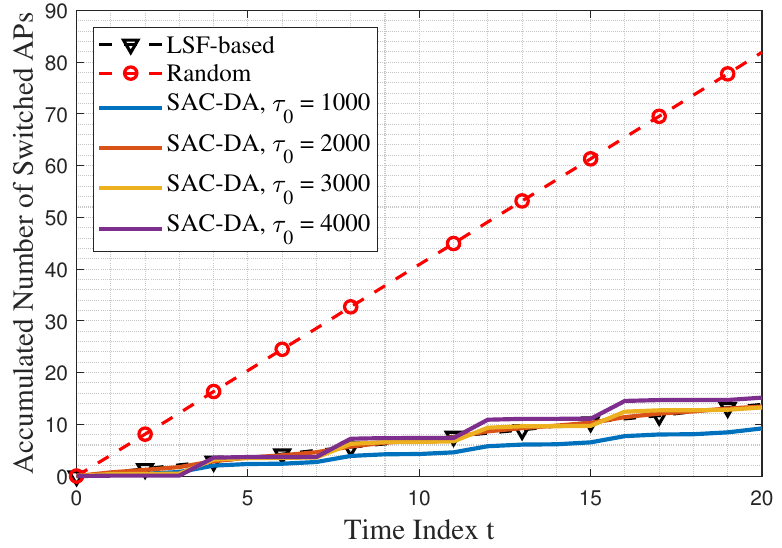}
			\vspace{-1.5em}
			\caption{Accumulated number of HOs.}
			\label{sfig:nOfHOs_DA_P_S}
		\end{subfigure}
		\vspace{-0.5em}
		\caption{Movement direction-assisted power-penalized system.}
		\label{fig:Penalty}
		\vspace{-1.5em}
	\end{minipage}
	\quad \quad \quad \quad
	\begin{minipage}{.4\textwidth}
		\centering
		\begin{subfigure}[t]{1\textwidth}
			\centering
			\includegraphics[width=1\textwidth]{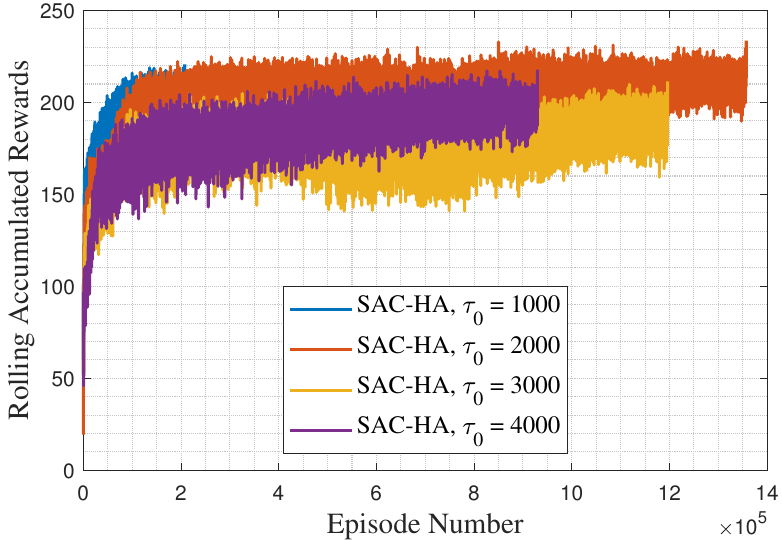}
			\vspace{-1.5em}
			\caption{Training.}
			\label{sfig:Training_HA_P_S}
		\end{subfigure}
		\\
		\begin{subfigure}[t]{1\textwidth}
			\centering
			\includegraphics[width=1\textwidth]{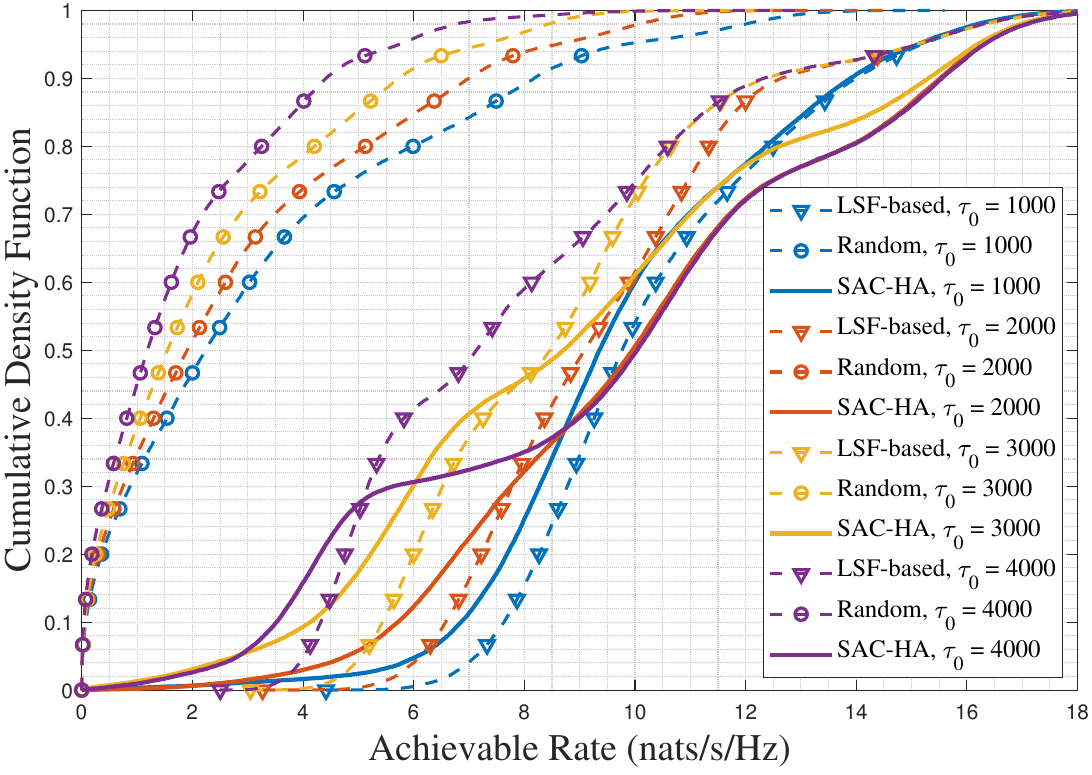}
			\vspace{-1.5em}
			\caption{CDF of achievable rate during the evaluation.}
			\label{sfig:CDF_HA_P_S}
		\end{subfigure}
		\\
		\begin{subfigure}[t]{1\textwidth}
			\centering
			\includegraphics[width=1\textwidth]{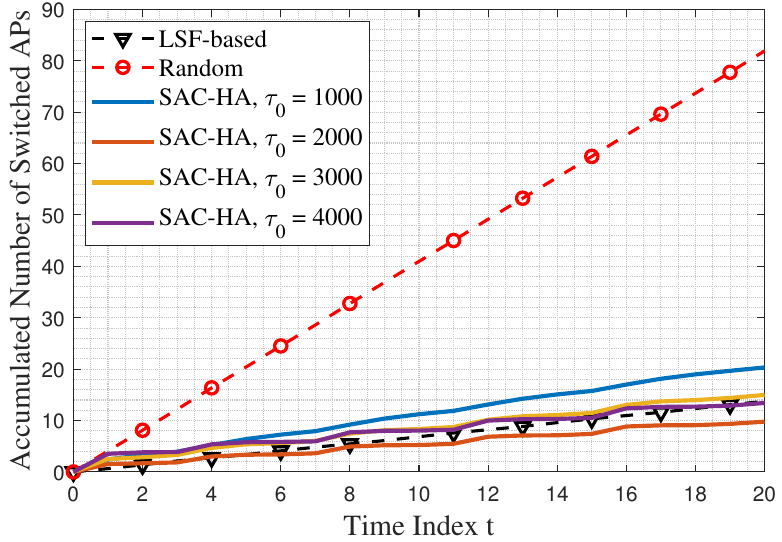}
			\vspace{-1.5em}
			\caption{Accumulated number of HOs.}
			\label{sfig:nOfHOs_HA_P_S}
		\end{subfigure}
		\vspace{-0.5em}
		\caption{History-assisted power-penalized system; $\beta_{\rm threshold} = {\rm PL}(\bar{d}_{bu}^{(t)} = 300~{\rm m})$.}
		\label{fig:Penalty_HA-P}
		\vspace{-1.5em}
	\end{minipage}
\end{figure*}


In Fig.~\ref{fig:noPenalty}, we test our formulation when no HO penalty exists and when the number of served users is set to one (all the APs have the same load). Using this setting, the LSF-based HO scheme will be the global optimum. In Fig.~\ref{fig:noPenalty}(\subref{sfig:noPenalty_Training_compiled}), we plot the rolling accumulated reward during training versus the number of episodes. The plots show both the reward using our framework for the continuous action space and using our framework but through a discrete action space (using dueling DDQN). The plots also show the number of episodes needed to reach the maximum rolling reward at which we save the actor DNN to use as our HO policy.

\begin{table}[t]
	\scriptsize
	\centering
	\caption{Size on disk and other info for the HO policy (Actor DNN); system specs: Windows~11 OS using NTFS file system and  Intel core i7-10750H CPU.}
	\vspace{-0.8em}
	\begin{tabular}{|p{0.15\linewidth}|p{0.16\linewidth}|p{0.14\linewidth}|p{0.16\linewidth}|p{0.14\linewidth}|}
		\hline
		\hline
		&\tiny \textit{\textbf{Dueling DDQN (discrete) $B = 9$, $B_{\rm con} = 5$ }} & 
		\tiny 
		\textit{\textbf{SAC (continuous) $B = 9$, $B_{\rm con} =$ any}} &
		\tiny 
		\textit{\textbf{Dueling DDQN (discrete) $B = 27$, $B_{\rm con} = 5$}} &
		\tiny 
		\textit{\textbf{SAC (continuous) $B = 27$, $B_{\rm con} =$ any}}\\
		\hline
		Size on disk (kbytes) & 103 & 31 & 32000  & 60
		\\
		\hline
		Size of action space & 126 & 9 & 80730 & 27
		\\
		\hline
		Policy response time (msec) & 0.474598 & 0.382671 & 1.473644 & 0.390981
		\\
		\hline
		\hline
	\end{tabular}
	\label{table:diskSpace}   
	\vspace{-2em}
\end{table}

In Figures~\ref{fig:noPenalty}(\subref{sfig:noPenalty_CDF_9DUs}) and~\ref{fig:noPenalty}(\subref{sfig:noPenalty_CDF_27DUs}), we plot the cumulative density function (CDF) of the achievable rate using our HO policy, where during this evaluation phase we test the trained actor DNN on new realizations. The results show that using a continuous action space achieves the global optimum. While using the discrete action space, we can only achieve the global optimum for a small network size, as seen in Fig.~\ref{fig:noPenalty}(\subref{sfig:noPenalty_CDF_27DUs}). Therefore, in contrast to a discrete action space, the utilization of a continuous action space offers an approach that adheres to the scalability description discussed in~\cite{9064545, surveyAmmar9650567}. This definition specifies that a solution for cell-free networks is scalable if it maintains a finite complexity and resource demands as the number of users or the network size grows.

The computational complexity for training DNNs with fully connected layers using the SAC algorithm can be summarized as $\mathcal{O}(T C ( L_{\rm actor} N_{\rm actor}^2 + 2 L_{\rm critic} N_{\rm critic}^2))$, where $T$ is the total number of training steps, $C$ is the batch size used for training which represents the computational cost per training step, $(L_{\rm actor}, L_{\rm critic})$ are the number of layers of the actor and critic DNNs, and $(N_{\rm actor}, N_{\rm critic})$ are the number of neurons in the (actor, critic) networks. The factor of 2 for the critic DNNs accounts for the two critic networks used in SAC. The square terms on $N_{\rm actor}$ and $N_{\rm critic}$ arise from the $N\times N$ connections between two layers each having size $N$. However, the DRL framework can train the DNN offline which means it should not be a concern for our proposed solution.

\begin{figure*}[t]
	\centering
	\begin{subfigure}[t]{0.4\textwidth}
		\centering
		\includegraphics[width=1\textwidth]{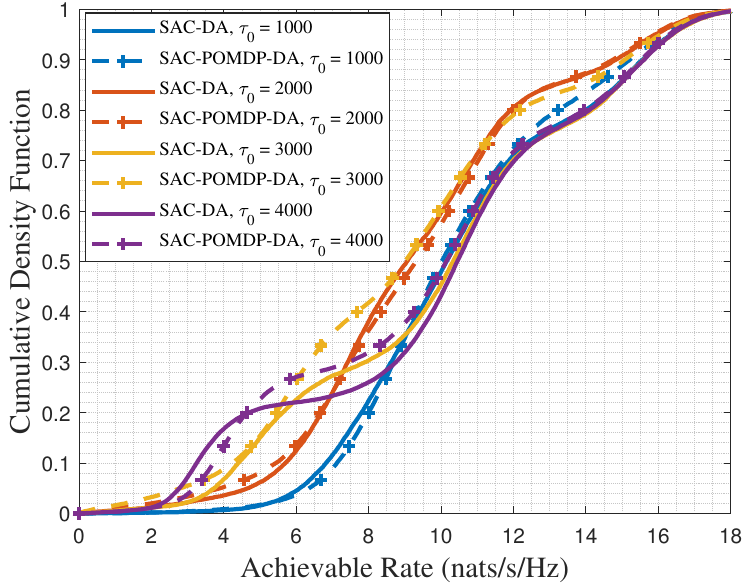}
		\vspace{-1.5em}
		\caption{CDF of achievable rate during the evaluation.}
		\label{sfig:CDF_DA_P_S_POMDP}
	\end{subfigure}
	$\quad \quad \quad \quad $
	\begin{subfigure}[t]{0.4\textwidth}
		\centering
		\includegraphics[width=1\textwidth]{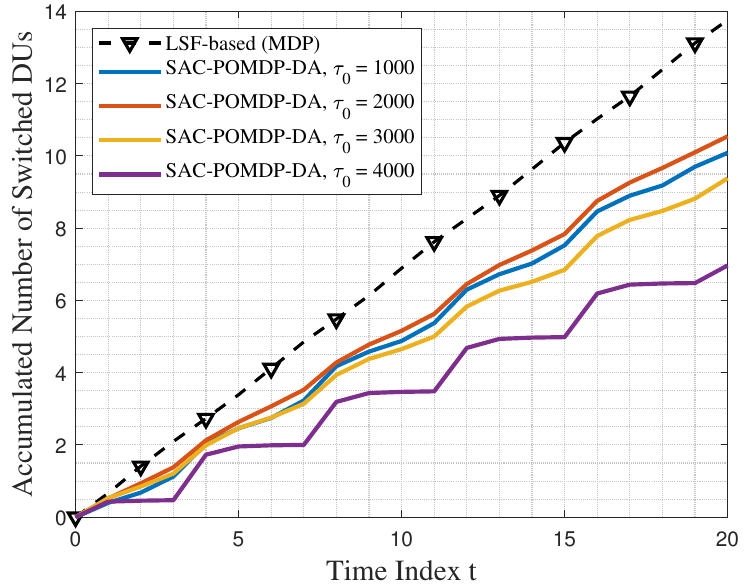}
		\vspace{-1.5em}
		\caption{Accumulated number of HOs}
		\label{sfig:nOfHOs_POMDP_DA_P_S}
	\end{subfigure}
	\vspace{-0.5em}
	\caption{Effect of partial observability (POMDP) on the Movement direction-assisted power-penalized system.}
	\label{fig:partialObserv}
	\vspace{-1em}
\end{figure*}

For the convergence of the training, we noticed that the rolling accumulated (sum) reward per episode, which is averaged over the last 100 episodes, can alternate within a specific range at the end of convergence. This is a common behavior when training using DRL. It is caused by many things such as the batch updates from the replay memory that contains both early and late experiences. Another cause is the stochastic policy used by SAC, where the agent samples actions from a probability distribution that injects variability in performance. To deal with this issue and ensure an efficient policy, we continuously save the actor DNN (the HO policy) whenever we achieve a new maximum rolling accumulated reward per episode. We therefore ensure that we have a copy of the best HO policy which will be used once training is finished. 

In Table~\ref{table:diskSpace}, we plot the size on disk, size of action space, and the policy response time for our HO policy. The policy response time is the average time needed to obtain the action after supplying the observation. For the continuous action space, we include in our calculation the time needed to obtain the connection decisions from the continuous action representation, i.e., including the delay of the mapping function in Fig.~\ref{fig:mob_NNoutput_conversion}. Our results show that our proposed approach which uses a vector of continuous action space is more scalable than employing a discrete action space. The results also show that our solution can provide a policy response time that is less than the duration of the 5G new radio NR subframe, which 
would allow us to deploy the policy for real-time execution.

In Figures~\ref{fig:Penalty} and~\ref{fig:Penalty_HA-P}, we plot the performance of the movement direction-assisted system (DA) and the history-assisted system (HA), respectively. In Figures~\ref{fig:Penalty}(\subref{sfig:Training_DA_P_S}) and~\ref{fig:Penalty_HA-P}(\subref{sfig:Training_HA_P_S}), we plot the rolling accumulated rewards of episodes during training for the two systems with a rolling window of size $100$~episodes. The figures show that the rolling reward increases with the number of episodes till it reaches some point where it starts to oscillate. As noted earlier, we save our HO policy (the actor DNN) whenever we have a new maximum rolling reward. Our results show that the DRL framework is effective in training a policy to maximize the reward function.

In Figures~\ref{fig:Penalty}(\subref{sfig:CDF_DA_P_S}) and~\ref{fig:Penalty_HA-P}(\subref{sfig:CDF_HA_P_S}), we plot the CDF of the achievable rate with HOs serving as a penalty using our HO policy represented through the trained actor DNN. The results show that at higher values for $\tau_0$, i.e., higher cost to initiate HOs, there is more space to outperform LSF-based association. Our results show about $27\%$ and $20\%$ increase in the data rate for the DA and the HA variants of the system, respectively. This highlights the importance of advanced HO strategies that can predict the serving set of the user.

In Figures~\ref{fig:Penalty}(\subref{sfig:nOfHOs_DA_P_S}) and~\ref{fig:Penalty_HA-P}(\subref{sfig:nOfHOs_HA_P_S}), we plot the accumulated number of switched APs, which is the same as the number of added APs to the serving set of the user. This number reflects the number of HO operations. In general, the DRL-based HO scheme produces a smaller or equal number of HOs compared to the LSF-based association, however, the DRL-based HO scheme seems to gather these HOs to make them occur concurrently. By initiating HOs less often, a lower cost for HOs is incurred.

The choice of using either the DA system or the HA system is highly dependent on the availability of the needed information. The DA system is very simple and only requires two consecutive locations for the user to construct the movement direction. On the other hand, the HA system does not require location tracking data, however, it needs a memory to store the history statistics of the LSF. Our results show that the DA system produces a slightly better results than the HA one. A possible reason for this is that the HA system is more complex and as shown in~\eqref{eq:zeta_history} requires choosing an LSF threshold $\beta_{\rm threshold}$ to classify the channel quality and a discount factor $\gamma_{\rm o}$ to prioritize newer history for the LSF.


\begin{figure}[t]
	\centering
	\includegraphics[width=0.8\columnwidth]{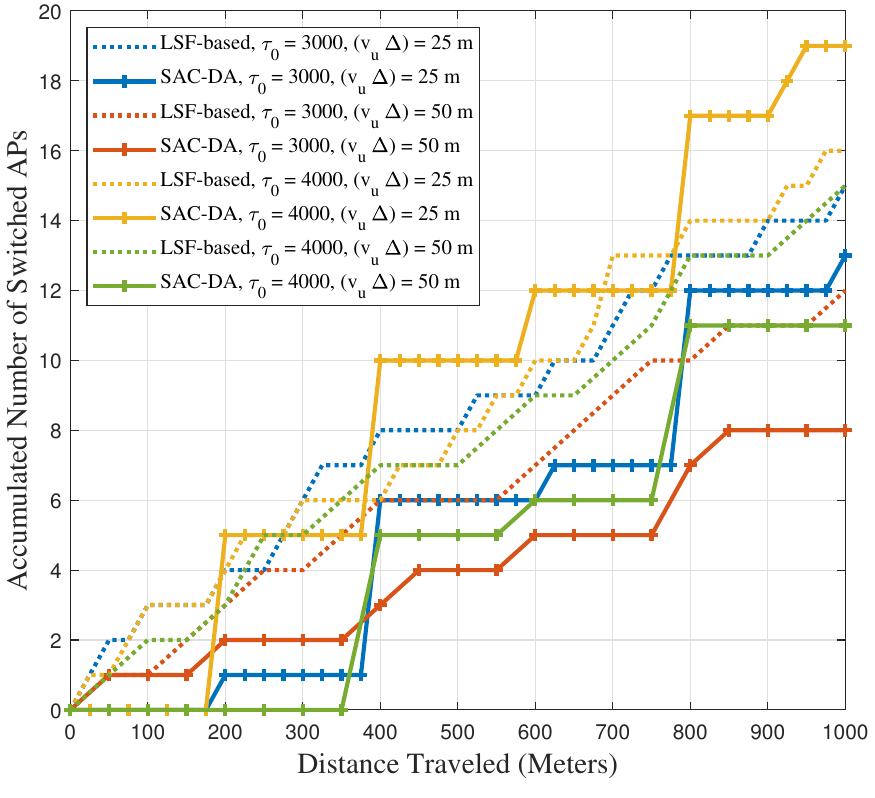}
	\vspace{-1em}
	\caption{Accumulated number of switched APs (number of HOs) from \textbf{random} network trials. For a fair comparison, same trials (plotted with the same color) are used for both DRL (SAC-DA) and LSF-based schemes.}
	\label{nOfHOs_diff_distResolution_tau0}
	\vspace{-0.7em}
\end{figure}

\subsection{Evaluation of DRL framework with Partial Observability}

In Fig.~\ref{fig:partialObserv}, we plot the performance of our DRL framework when the states are partially observable (as explained in Section~\ref{sec:POMDP}). In this simulation, the DRL model is based on a POMDP instead of an MDP. Notably, the LSF-based association cannot be implemented because we cannot observe the channels between the user and all the APs in the network. Our results demonstrate that the actor DNN can still be trained to make optimized HO decisions even under POMDP conditions. Our simulations reveal that, for some settings assuming partial observability, we experience negligible losses. Specifically, we observe losses of 
$[3\%, 10\%, \sim0\%, 1\%]$ in the achievable rate for $\tau_0 = [4000, 3000, 2000, 1000]$. Additionally, the number of HOs seems to remain low compared to an LSF-based approach that is based on an MDP, which indicates that DRL can still minimize the number of HOs under partial observability. This outcome emphasizes that DRL has the potential to implement advanced HO policies in imperfect situations.

As a final test, Fig.~\ref{nOfHOs_diff_distResolution_tau0} compares the accumulated number of HOs from using DRL versus the standard LSF scheme, under different time durations $\bar{\Delta}$ for HO decision checks. The results indicate that DRL is able to learn a policy that concentrates HOs into specific time slots, effectively reducing the overall HO frequency. This behavior helps lower the signaling overhead and resource cost associated with intiating the HOs over sparse time slots. Moreover, in most scenarios, DRL results in fewer total HOs compared to the LSF-based approach, highlighting its ability to optimize HO timing and reduce unnecessary transitions in serving APs. In some cases, such as ($\tau_0=4000$, $\left(\mathtt{v}_u \bar{\Delta}\right) = 25{\rm m}$) shown in Fig.~\ref{nOfHOs_diff_distResolution_tau0}, LSF could result in a lower total number of HOs at some points, however, this does not contradict with the behaviour designed by the objective function in~\eqref{eq:reward_HOPenalty}. This behaviour prioritizes lowering the frequency of HOs and hence lowering the overhead for initiating HOs in multiple time slots instead of few.

\subsection{Deployment in Real-World}\label{section:DeploymentLocation}
Our DRL solution allows the network and users to self-learn the best HO decisions in the UC-mMIMO network. The network operators have the freedom to tune different parameters such as the HO penalty parameter~$\tau_0$, which is the basic overhead for initiating HOs, to control the HO behavior. The DRL framework can be trained offline until it achieves acceptable behavior, then the trained DNN (the policy) can be implemented in real time as measured in Table~\ref{table:diskSpace}.

It is worth mentioning that the HO policy, represented by the actor DNN, can still be improved even during the deployment phase through online training. This is one of the powerful features of DRL. Hence, in an actual deployment, mobile operators can label their trained policies, and whenever they have a better one they can deploy it to replace an older version. In case of dissatisfaction or undesired behavior, which can happen in AI models, mobile operators can always roll back to older versions of HO policies.

The main challenge in deploying our DRL framework is gathering the needed network information at either the network side or the user side. This depends on the deployed location of the the DRL framework as discussed in Section~\ref{section:ActorCriticLoc}. Gathering this information is feasible and it is only restricted to the observation and reward. However, the support of POMDPs and partial observability can relax this requirement. There are also more advanced systems that can be developed as we state in the conclusions as future work.


\section{Conclusion}\label{sec:conclusion}
We have developed a DRL-based solution that manages HO decisions in the UC-mMIMO network scheme. Our solution trains a DNN to serve as the HO policy for mobile users. To assist in HO decisions, our system comprises two variants: the first utilizes movement-direction assisted observations, and the second employs history-assisted observations. Both systems use unique information to aid in HO decision-making and minimize the number of HOs required. In comparison to LSF-based HOs, our DRL-based HO policy obtained for the movement direction-assisted and history-assisted variants can improve the achievable rate by approximately $27\%$ and $20\%$, respectively. This improvement is achieved by intelligently combining HO operation within the same time indices. Crucially, to enable scalability, we use a continuous action space, which is more scalable than discrete action space approach, in which the policy response time and the disk size for the DNN is reduced by $75\%$ and $98\%$, respectively. 
Furthermore, our scheme supports real-time control with latency lower than $0.4~{\rm ms}$ and can operate under partial observability.

A future direction for our study is to consider jointly decreasing the number of HOs for the network users. This approach should ensure that both the action space and state space do not grow exponentially with the number of users, otherwise, DRL will not be able to train the DNN to serve as an effective joint HO policy. A crucial aspect of this future topic is to build a solution that prioritizes both scalability and reusability.
%
%

\appendix
\section{}
\subsection{Derivations for~\eqref{eq:rate_LB}: Lower Bound for Channel Capacity}\label{Appendix:rate_LB}
The derivations, presented herein, follow the same steps as those in~\cite{POMDP_J} with some minor differences. The power of the desired signal can be derived as
\allowdisplaybreaks{
	\begin{align}
		&\xi_{1,u}[n]
		=
		\left| \mathbb{E}\left\{{\rm DS}_u[n]\right\} \right|^2
		\nonumber \\[-5pt]
		& =
		\Big|
		\rho_{u}[n - n_{\rm est}]
		\sum_{b \in \mathcal{C}_u} \sqrt{\eta_{bu}} 
		\mathbb{E}\left\{
		{\bf h}_{bu}^T[n_{\rm est}]
		{\bf \hat{h}}_{bu}^*
		\right\}
		\Big|^2
		\nonumber\\[-5pt]
		&=
		\Big|
		\rho_{u}[n - n_{\rm est}]
		\sum_{b \in \mathcal{C}_u} \sqrt{\eta_{bu}}
		\nonumber\\[-5pt]
		& \quad \quad \quad \quad \quad \quad \quad
		\times
		\mathbb{E}\left\{
		(\mathrm{ {\bf e}}_{bu}^T[n_{\rm est}] + {\bf \hat{h}}_{bu}^T[n_{\rm est}])
		{\bf \hat{h}}_{bu}^*[n_{\rm est}]
		\right\}
		\Big|^2
		\nonumber\\[-5pt]
		&
		=
		M^2
		\rho_{u}^2[n - n_{\rm est}]
		\Big|
		\sum_{b \in \mathcal{C}_u} \sqrt{\eta_{bu}} 
		\psi_{bu}
		\Big|^2
\end{align}}
$\!\!\!$where $\mathbb{E}\left\{
\mathrm{ {\bf e}}_{bu}^T[n_{\rm est}]
{\bf \hat{h}}_{bu}^*[n_{\rm est}]
\right\} = 0$ due to the uncorrelated terms, and $\mathbb{E}\left\{
{\bf \hat{h}}_{bu}^T[n_{\rm est}]
{\bf \hat{h}}_{bu}^*[n_{\rm est}]
\right\} = M \psi_{bu}$ leading to $\mathbb{E}\left\{
{\bf h}_{bu}^T[n_{\rm est}]
{\bf \hat{h}}_{bu}^*[n_{\rm est}]
\right\} = M \psi_{bu}$.

The beamformer uncertainty can be derived as
\begin{align}\label{eq:BU_analysis}
	&\xi_{2,u}[n]
	= 
	\mathbb{E}\left\{\left|{\rm BU}_u[n]\right|^2\right\} 
	\nonumber \\[-5pt]
	& =
	\mathbb{E}\bigg\{\bigg|
	\rho_{u}[n - n_{\rm est}]\sum_{b \in \mathcal{C}_u} \sqrt{\eta_{bu}}
	\nonumber\\[-5pt]
	&
	\quad \quad \quad
	\times
	\left(
	{\bf h}_{bu}^T[n_{\rm est}]
	{\bf \hat{h}}_{bu}^*[n_{\rm est}]
	-
	\mathbb{E}\left\{
	{\bf h}_{bu}^T[n_{\rm est}]
	{\bf \hat{h}}_{bu}^*[n_{\rm est}]
	\right\}
	\right)
	\bigg|^2\bigg\}
	\nonumber \\[-5pt]
	& \stackrel{(a)}{=}
	\rho_{u}^2[n - n_{\rm est}]
	\sum_{b \in \mathcal{C}_u} 
	\eta_{bu}
	\nonumber\\[-5pt]
	&
	\ 
	\times  
	\left(
	\mathbb{E}\left\{ \left| {\bf h}_{bu}^T[n_{\rm est}]
	{\bf \hat{h}}_{bu}^*[n_{\rm est}]
	\right|^2
	\right\}
	-
	\left|
	\mathbb{E}\left\{
	{\bf h}_{bu}^T[n_{\rm est}]
	{\bf \hat{h}}_{bu}^*[n_{\rm est}]
	\right\}\right|^2
	\right)
\end{align}
where $(a)$ follows because the variance of a sum of independent random variables equals the sum of the variances. The first term in~\eqref{eq:BU_analysis} can be calculated as
\begin{align}\label{eq:BU_analysis2}
	&\mathbb{E}\left\{ \left| {\bf h}_{bu}^T[n_{\rm est}]
	{\bf \hat{h}}_{bu}^*[n_{\rm est}]
	\right|^2
	\right\}
	=
	\nonumber\\ 
	& \quad \quad \quad
	\mathbb{E}\left\{ \left| \mathrm{ {\bf e}}^T[n_{\rm est}]
	{\bf \hat{h}}_{bu}^*[n_{\rm est}]
	\right|^2
	\right\}
	+ \mathbb{E}\left\{ \left\|
	{\bf \hat{h}}_{bu}^*[n_{\rm est}]
	\right\|^4
	\right\}
	\nonumber \\[-5pt]
	&=
	\left(M^2  \psi_{bu} \left(\beta_{bu} - \psi_{bu} \right) \right)
	+
	2 M^2 \psi_{bu}^2
\end{align}
The second term in~\eqref{eq:BU_analysis2} follows from the fact that for a Gaussian random variable (RV) $X\sim \mathcal{N}(0, \psi_{bu})$ the RV $Y = X^2$ has a chi-squared distribution. 
Then,
\begin{align}
	\xi_{2,u}[n]
	& =
	M^2 \rho_{u}^2[n - n_{\rm est}]
	\sum_{b \in \mathcal{C}_u} 
	\eta_{bu} 
	\beta_{bu}\psi_{bu}
\end{align}
The channel aging term can be calculated as
\begin{align}
	&\xi_{3,u}[n]
	=
	\mathbb{E}\left\{\left| {\rm CA}_u[n] \right|^2\right\} 
	\nonumber \\[-5pt]
	& =
	\bar{\rho}_{u}^2[n - n_{\rm est}]
	\mathbb{E}\left\{ \left|
	\sum_{b \in \mathcal{C}_u} \sqrt{\eta_{bu} \beta_{bu}} 
	{\bf v}_{bu}^T[n]
	{\bf \hat{h}}_{bu}^*[n_{\rm est}]
	\right|^2\right\}
	\nonumber \\[-5pt]
	& =
	\bar{\rho}_{u}^2[n - n_{\rm est}]
	\left(
	\sum_{b \in \mathcal{C}_u} \eta_{bu} \beta_{bu} 
	\Xi_{1,bu}
	\right.
	\nonumber \\[-5pt]
	& \quad \quad
	+
	\left.
	\sum_{b \in \mathcal{C}_u}
	\sum_{b' \in \mathcal{C}_u, b' \ne b} 
	\sqrt{\eta_{bu} \beta_{bu}}
	\sqrt{\eta_{b'u} \beta_{b'u}}
	\Xi_{2,bb'u}
	\right)
\end{align}
with
\begin{align}
	\Xi_{1, bu} &= \mathbb{E}\left\{ \left|
	{\bf v}_{bu}^T[n]
	{\bf \hat{h}}_{bu}^*[n_{\rm est}]
	\right|^2\right\}
	= M^2 \psi_{bu}
	\\[-5pt]
	\Xi_{2,bb'u} & =
	\mathbb{E}\left\{
	\left(
	{\bf v}_{bu}^T[n]
	{\bf \hat{h}}_{bu}^*[n_{\rm est}]\right)^*
	\left(
	{\bf v}_{b'u}^T[n]
	{\bf \hat{h}}_{b'u}^*[n_{\rm est}]\right)
	\right\} = 0
\end{align}
\begin{align}
\text{Then,}\quad 
\xi_{3,u}[n]
	=
	M^2 \bar{\rho}_{u}^2[n - n_{\rm est}]
	\sum_{b \in \mathcal{C}_u} \eta_{bu} \beta_{bu} 
	\psi_{bu}
\end{align}
Using the definition of $\bar{\rho}_{u}[\cdot]$, we can combine $\xi_{2,u}$ and $\xi_{3,u}$ into a single expression as
\begin{align}
	\xi_{2,3,u}[n] & =
	\xi_{2,u}[n] + \xi_{3,u}[n]
	=
	M^2
	\sum_{b \in \mathcal{C}_u} \eta_{bu} \beta_{bu} 
	\psi_{bu}
\end{align}
The interference caused by the APs due to serving user $u'$ can be calculated as
\begin{align}
	&\xi_{4,uu'}[n]
	=
	\mathbb{E}\left\{\left| {\rm MI}_{uu'}[n] \right|^2\right\} 
	\nonumber \\[-5pt]
	& = \mathbb{E}\Big\{\Big| \sum_{b' \in \mathcal{C}_{u'}} \sqrt{\eta_{b'u'}} {\bf h}_{b'u}^T[n] \hat{\bf h}_{b'u'}^{*}[n_{\rm est}] \Big|^2\Big\}
	\nonumber \\[-3pt]
	& =
	\sum_{b' \in \mathcal{C}_{u'}} \eta_{b'u'} \Xi_{3,b'uu'}
	\nonumber \\[-3pt]
	& \quad \quad
	+
	\sum_{b' \in \mathcal{C}_{u'}} \sum_{b'' \in \mathcal{C}_{u'}, b'' \ne b'} \sqrt{\eta_{b'u'}} \sqrt{\eta_{b''u'}} \Xi_{4,b'b''uu'}
\end{align}
where
\begin{align}
	&\Xi_{3,b'uu'} = \mathbb{E}\left\{\left| {\bf h}_{b'u}^T[n] \hat{\bf h}_{b'u'}^{*}[n_{\rm est}] \right|^2\right\}
	\nonumber \\
	& \stackrel{(a)}{=}
	\rho_{u}^2[n - n_{\rm est}] \mathbb{E}\left\{\left| {\bf h}_{b'u}^T[n_{\rm est}] \hat{\bf h}_{b'u'}^{*}[n_{\rm est}] \right|^2\right\}
	\nonumber \\
	& \quad \quad \quad
	+
	\beta_{b'u}
	\bar{\rho}_{u}^2[n - n_{\rm est}]
	M^2 
	\psi_{b'u'}
	\nonumber \\
	& \stackrel{(b)}{=}
	\begin{cases}
		\begin{aligned}
		&
		\rho_{u}^2[n - n_{\rm est}] \left( M^2\beta_{b'u}\psi_{b'u'} + M^2 \psi_{b'u} \psi_{b'u'} \right)
		\nonumber \\
		& \quad \quad \quad
		+
		\beta_{b'u}
		\bar{\rho}_{u}^2[n - n_{\rm est}]
		M^2 
		\psi_{b'u'},
		\end{aligned}
		& \text{if}\ u' \in \mathcal{U}_i
		\\
		\begin{aligned}
		&
		\rho_{u}^2[n - n_{\rm est}] \left( M^2\beta_{b'u}\psi_{b'u'} \right)
		\nonumber \\
		& \quad \quad \quad
		+
		\beta_{b'u}
		\bar{\rho}_{u}^2[n - n_{\rm est}]
		M^2 
		\psi_{b'u'},
		\end{aligned}
		& \text{if}\ u' \notin \mathcal{U}_i
	\end{cases}
	\nonumber \\
	& \stackrel{(c)}{=}
	\begin{cases}
		M^2\beta_{b'u}\psi_{b'u'} + M^2 \rho_{u}^2[n - n_{\rm est}] \psi_{b'u} \psi_{b'u'}
		,& \text{if}\ u' \in \mathcal{U}_i
		\\
		M^2 \beta_{b'u}\psi_{b'u'}
		,& \text{if}\ u' \notin \mathcal{U}_i
	\end{cases}
\end{align}
where $(a)$ is obtained using~\eqref{eq:channelEvolution_data}, $(b)$ follows from the expression of the estimated channel ${\bf \hat{h}}_{bu}$ in~\eqref{eq:est_chan} where user $u$ is assumed using the pilot $i$, i.e., $u \in \mathcal{U}_i$, and $(c)$ follows using the definition $\bar{\rho}_{u}[n - n_{\rm est}] = \sqrt{ 1 - |\rho_{u}[n - n_{\rm est}]|^2 }$. Also,
\begin{align}	
	&\Xi_{4, b'b''uu'} =
	\mathbb{E}\left\{\left( {\bf h}_{b'u}^T[n] \hat{\bf h}_{b'u'}^{*}[n_{\rm est}] \right)^* \left( {\bf h}_{b''u}^T[n] \hat{\bf h}_{b''u'}^{*}[n_{\rm est}] \right)\right\}
	\nonumber \\
	&=
	\resizebox{0.95\linewidth}{!}
	{$\displaystyle
	\rho_{u}^2[n - n_{\rm est}] \mathbb{E}\left\{ {\bf h}_{b'u}^T[n_{\rm est}] \hat{\bf h}_{b'u'}^{*}[n_{\rm est}] \right\}
	\mathbb{E}\left\{
	{\bf h}_{b''u}^T[n_{\rm est}] \hat{\bf h}_{b''u'}^{*}[n_{\rm est}] \right\}
	$}
	\nonumber\\
	&=
	\begin{cases}
		\resizebox{0.7\linewidth}{!}
		{$\displaystyle
		\rho_{u}^2[n - n_{\rm est}] \left( M \sqrt{\psi_{b'u} \psi_{b'u'} } \right)
		\left( M \sqrt{\psi_{b''u} \psi_{b''u'} } \right)
		$}
		,& \text{if}\ u' \in \mathcal{U}_i
		\\[-5pt]
		0,& \text{if}\ u' \notin \mathcal{U}_i
	\end{cases}
\end{align}
Then,
\vspace{-0.5em}
	\begin{align}
		&\xi_{4,uu'}[n]
		=
		\nonumber\\
		&
		\begin{cases}
			\begin{aligned}
				&
				M^2 \big(\sum_{b' \in \mathcal{C}_{u'}} \eta_{b'u'}
				\beta_{b'u}\psi_{b'u'}
				+
				\rho_{u}^2[n - n_{\rm est}]
				\\
				&\quad \quad 
				\times
				\big|\sum_{b' \in \mathcal{C}_{u'}} \sqrt{\eta_{b'u'}} \sqrt{\psi_{b'u} \psi_{b'u'} }
				\big|^2
				\big)
			\end{aligned}
			,& \text{if}\ u' \in \mathcal{U}_i
			\\
			M^2 \sum_{b' \in \mathcal{C}_{u'}} \eta_{b'u'}
			\beta_{b'u}\psi_{b'u'}
			,& \text{if}\ u' \notin \mathcal{U}_i
		\end{cases}
\end{align}
Then, the expression in~\eqref{eq:rate_LB} for the achievable rate is obtained through vector multiplication to represent the summation.


%
%
%
%
%
%

%
\footnotesize
\bibliography{Mob_ML_References}
\bibliographystyle{ieeetr}

\end{document}